\documentclass[showpacs,prl,reprint]{revtex4-1}

\usepackage{hyperref}
\usepackage{graphicx}
\usepackage{color}
\usepackage{amsbsy}
\usepackage{pstricks}
\usepackage{multirow}
\usepackage{amsmath}
\usepackage{lipsum}

\begin{document}

\title{Atomic-Scale Magnetometry of Dynamic Magnetization}

\date{\today}

\author{J. van Bree}
\email{j.v.bree@tue.nl}
\author{M. E. Flatt{\'e}}
\email{michael\_flatte@mailaps.org}
\affiliation{Department of Physics and Astronomy and Optical Science and Technology Center, University of Iowa, Iowa City, Iowa 52242, USA}


\begin{abstract}
The spatial resolution of imaging magnetometers has benefited from scanning probe techniques. The requirement that the sample perturbs the scanning probe through a magnetic field external to its volume limits magnetometry to samples with pre-existing magnetization. We propose a magnetometer in which the perturbation is reversed: the probe's magnetic field generates a response of the sample, which acts back on the probe and changes its energy. For an NV$^-$~spin center in diamond this perturbation changes the fine-structure splitting of the spin ground state. Sensitive measurement techniques using coherent detection schemes then permit detection of the magnetic response of paramagnetic and diamagnetic materials. This technique can measure the thickness of magnetically dead layers with better than $0.1$~\AA~accuracy. 
\end{abstract}

\maketitle

\begin{figure}[b]
\includegraphics[width=\columnwidth]{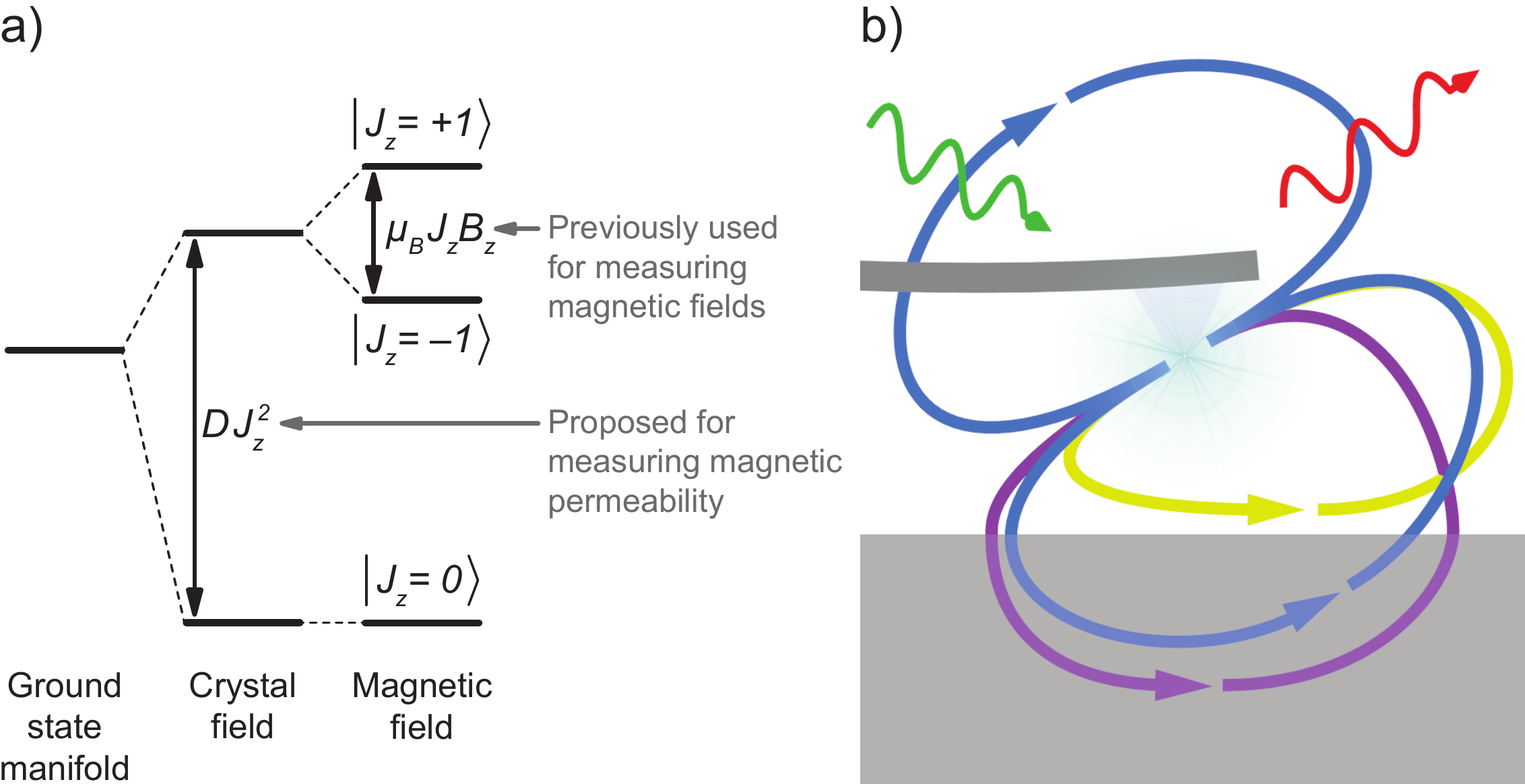}
\caption{(a) The NV$^-$ center's ground state spin $J=1$ is split by the crystal field and magnetic field. Conventional NV$^-$ magnetometry utilizes the splitting of the $|J_z = \pm 1\rangle$ states. We propose a way to measure magnetic response of materials using  the splitting between the $|J_z=0\rangle$ and $|J_z=\pm1\rangle$ states. (b) Implementation of magnetic-energy magnetometry. The spin of an NV$^-$ center is located at the apex of a scanning probe tip, optically initialized (green) and detected (red). The spin's magnetic induction (blue) is perturbed by the presence of the sample (gray), leading to modifications of the magnetic induction (diamagnetic, yellow; paramagnetic, purple).}
\label{fig:cartoon}
\end{figure}

Imaging of magnetic moments and magnetic fields advances a wide range of fields: nuclear magnetic resonance~\cite{Rabi1938} clarifies the structure of molecules and biological enzymes, superconducting quantum interference device magnetometry~\cite{Jaklevic1964} characterizes magnetically engineered multilayers, and magnetic resonance imaging (MRI)~\cite{Wehrli1992} distinguishes various types of tissue in medicine and biology. The spatial resolution of imaging magnetometers suffices, in principle, to observe interesting processes, such as biological activity in a cell, which are obscured from optical measurements by the diffraction limit~\cite{Balasubramanian2008}. In practice, however, the spatial resolution of even specialized MRI rarely surpasses $\mu$m~\cite{Ciobanu2002}, limited by the sensitivity at which the nuclear spins can be detected~\cite{Glover2002}. Various scanning probe techniques~\cite{Martin1987,Chang1992,Degen2009} improve this spatial resolution. A promising approach, NV$^-$-center magnetometry~\cite{Hong2013}, uses a defect formed by a substitutional nitrogen atom and adjacent vacancy site in a diamond crystal. The long spin-coherence time of this defect allows optical initialization and detection, and coherent manipulation with microwaves~\cite{Oort1988,Childress2006}, resulting in exceptional magnetic field sensitivity and spatial resolution at ambient conditions~\cite{Balasubramanian2008,Maze2008}. These scanning-probe-based magnetometers require the sample's magnetic field to perturb the magnetically sensitive probe nearby. In NV$^-$-center-based magnetometry, for example, measurements of the splitting between the spin ground state $|J_z=\pm1\rangle$ states detect this magnetic field, see Fig.~\ref{fig:cartoon}(a). This scheme, however, requires the sample to possess an substantial magnetic field external to its volume, which excludes weak-moment films, as well as paramagnetic and diamagnetic materials, which lack such external magnetic fields in isolation.

Here we propose to overcome this disadvantage, by using the probe's magnetic field to perturb the sample instead of relying on the sample's magnetic field to perturb the probe. For any sample magnetic permeability differing from that of vacuum, the magnetic field of the probe will be dynamically altered, changing the magnetic energy stored in the probe's magnetic field. For this approach, depicted in  Fig.~\ref{fig:cartoon}(b), we predict that for an NV$^-$ center these changes in magnetic energy effectively translate into a modification of the crystal field splitting of the NV$^-$ center's spin ground state, see Fig.~\ref{fig:cartoon}(a). Techniques have already been developed to measure small changes in this splitting for thermometry purposes~\cite{Toyli2013,Kucsko2013,Neumann2013}. Our calculations show that the magnetic energy approach to NV$^-$-center magnetometry makes it possible to measure the magnetic permeabilities of diamagnetic and paramagnetic materials. For a unique application of this technique, we propose measuring the thickness of magnetically dead layers~\cite{Sun1999}. We show it is possible to determine this thickness with an accuracy superior to $0.1$~\AA~for experimentally realistic conditions.

\begin{figure*}
\includegraphics[width=0.95\textwidth]{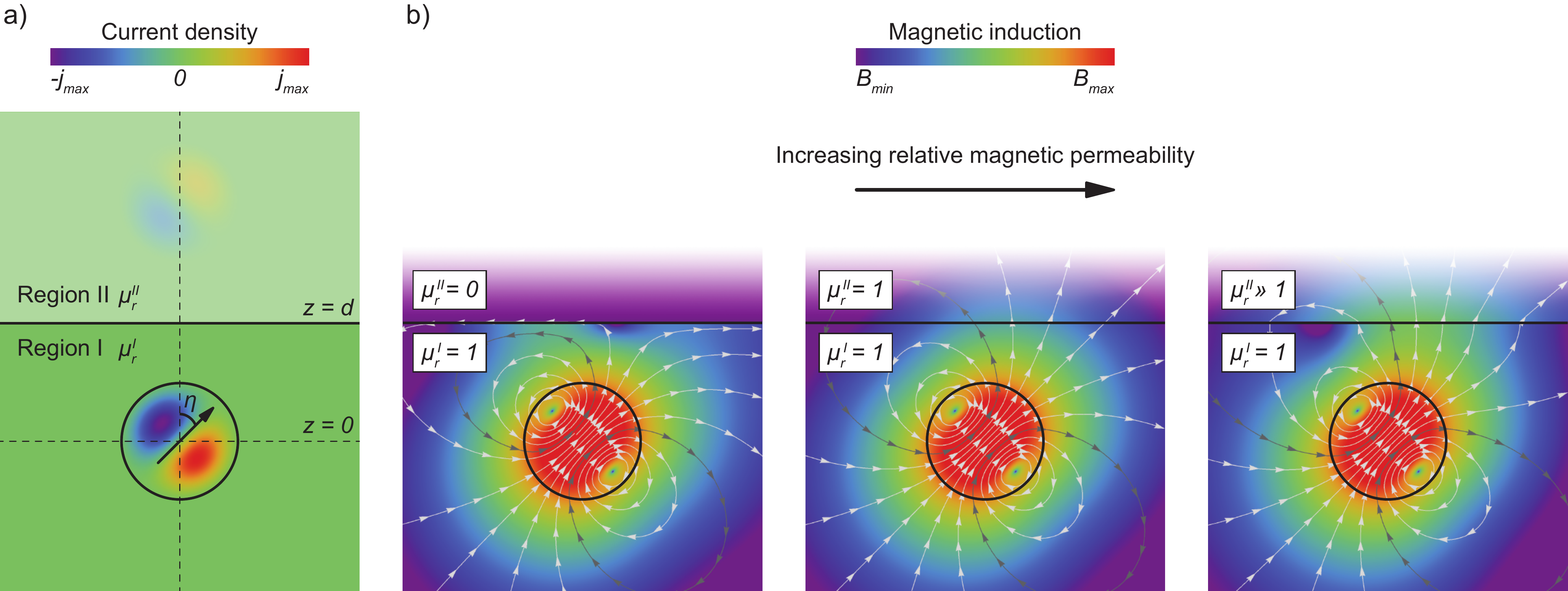}
\caption{(a) A spin is placed in region I with relative magnetic permeability $\mu_r^{\text{I}}$, adjacent to a semi-infinite region II with $\mu_r^{\text{II}}$, filling the half-space $z>d$. The spin makes an angle $\eta$ with respect to the normal of the interface between the two regions. The color indicates the magnitude of the (image) current distribution for ${\cal P}(r)=\left[j_0(\pi r/R)\right]^2$, a spherical Bessel function of zeroth order, and $\mu_r^{\text{II}}=2$. (b) The calculated magnetic induction for the situation as described in (a). The magnetic induction near the interface is parallel to the interface for diamagnetic substances ($\mu_r^{\text{II}}=0$, left), though perpendicular to the interface for paramagnetic materials ($\mu_r^{\text{II}}\gg1$, right). The magnitude of the magnetic induction is indicated by color, and its direction by the arrows of the streamlines.}
\label{fig:Bfield}
\end{figure*}

Consider a spin with total angular momentum $J$ in region I, placed in close proximity to region II with a different magnetic permeability $\mu_r$ [see Fig.~\ref{fig:Bfield}(a)]. In the absence of spin-orbit coupling, its magnetic moment density
\begin{align}
\langle{\boldsymbol \mu}({\bf x})\rangle = \frac{2\mu_B}{\hbar} \langle {\bf J}({\bf x}) \rangle = \frac{2\mu_B}{\hbar} \langle {\bf J} \rangle {\cal P}({\bf x}), \label{eq:mu}
\end{align}
depends on its probability density ${\cal P}({\bf x})$, and the expectation value of the spin operator ${\bf J}=(J_x,J_y,J_z)$; here $\mu_B$ is the Bohr magneton, and we took the $g$ factor to be $2$. In the Supplemental Material~\cite{SM} we show that this relation holds for any $N$-particle state, e.g., the complicated ground state of the NV$^-$ center comprising 6 electrons~\cite{Doherty2013}. To simplify the calculation, we now treat the interaction of the spin's magnetic moment with its environment classically; we will address its quantum-mechanical nature later on. The presence of a magnetic moment density requires a current density ${\bf j}({\bf x}) = \nabla \times \langle{\boldsymbol \mu}({\bf x})\rangle$, which provides a direct expression for  calculating, in the Coulomb gauge, the energy stored in a magnetic field~\cite{Jackson1998},
\begin{align}
E_{\text{mag}} = \frac{1}{2} \int {\bf j}({\bf x}) \cdot {\bf A}({\bf x})~d^3x. \label{eq:Emag}
\end{align}
Here ${\bf A}({\bf x})$ is the vector potential produced by ${\bf j}({\bf x})$. If ${\bf j}({\bf x})=0$ in region II,  Eq.~(\ref{eq:Emag}) determines the magnetic energy from  ${\bf A}({\bf x})$ in region I alone. The effect of region II on the spin's vector potential in region I can be included by replacing region II with an image current density~\cite{Jackson1998}
\begin{align}
\widetilde{{\bf j}}({\bf x}) = \frac{\mu_r^{\text{II}}-\mu_r^{\text{I}}}{\mu_r^{\text{II}}+\mu_r^{\text{I}}} \left(
\begin{array}{r}
 j_x(x,y,2d-z) \\
 j_y(x,y,2d-z) \\
-j_z(x,y,2d-z) \\
\end{array}\right)_{\hat{\bf x},\hat{\bf y},\hat{\bf z}},
\end{align} 
where for simplicity we neglect any surface current at the interface between the regions. A treatment of surface currents would be required for conductive materials with a nonzero component of their magnetization parallel to the surface normal at the interface of the two regions. The image current generates a vector potential $\widetilde{{\bf A}}({\bf x})$; the total vector potential in region I is then ${\bf A}({\bf x})+\widetilde{{\bf A}}({\bf x})$.

For ease of calculation we assume isolated spin systems are approximately spherically symmetric and limited to a sphere with radius $R < d$. We will show later on that is a fair approximation for the NV$^-$ center, even though that spin center has $C_{3v}$ symmetry~\cite{Doherty2013}. For a spin oriented such that its integrated magnetic moment makes an angle $\eta$ with respect to the $z$ axis [see Fig.~\ref{fig:Bfield}(a)], the spin's current density
\begin{align}
{\bf j}({\bf x}) &= 2\mu_B J \frac{d{\cal P}(r)}{dr} \notag\\
&\quad\times \left(
\begin{array}{c}
 0 \\
 \sin \eta \sin \phi \\
 \sin \eta \cos \theta \cos \phi - \cos \eta \sin \theta
\end{array}
\right)_{\hat{\bf r},\hat{\boldsymbol \theta},\hat{\boldsymbol \phi}},
\end{align}
for $r\leq R$. The vector potential resulting from this current distribution, calculated by expanding the Green's function in spherical harmonics and performing several (partial) integrations, is 
\begin{align}
{\bf A}({\bf x}) &= -\frac{2\mu_0 \mu_r^{\text{I}} \mu_B J}{r^2} \left(\int_0^r {\cal P}(r') r'^2 dr' \right) \notag \\
&\quad\times \left(
\begin{array}{c}
 0 \\
 \sin \eta \sin \phi \\
 \sin \eta \cos \theta \cos \phi - \cos \eta \sin \theta
\end{array}
\right)_{\hat{\bf r},\hat{\boldsymbol \theta},\hat{\boldsymbol \phi}},
\end{align}
for $r\leq R$, and the image current produces a vector potential
\begin{align}
\widetilde{{\bf A}}({\bf x}) &= \frac{\mu_r^{\text{II}}-\mu_r^{\text{I}}}{\mu_r^{\text{II}}+\mu_r^{\text{I}}} \left( \frac{\mu_0 \mu_r^{\text{I}} \mu_B J}{2 \pi \left(4 d^2-4 r d\cos \theta +r^2\right)^{3/2}} \right) \notag \\
&\hspace{-8mm}\times \left(
\begin{array}{c}
 - 2 d \sin \eta \sin \theta \sin \phi \\
 (r - 2 d \cos \theta) \sin \eta \sin \phi \\
 (r \cos \theta -2 d) \sin \eta \cos \phi + r \cos \eta \sin \theta
\end{array}
\right)_{\hat{\bf r},\hat{\boldsymbol \theta},\hat{\boldsymbol \phi}},
\end{align}
for $z<d$. These vector potentials determine the spin's magnetic induction ${\bf B}({\bf x}) = \nabla \times {\bf A}({\bf x})$, see Fig.~\ref{fig:Bfield}(b). The magnetic induction is either repelled from (drawn to) region II if $\mu_r^{\text{II}} < \mu_r^{\text{I}}$ ($\mu_r^{\text{II}} > \mu_r^{\text{I}}$), as the magnetization in region II induced by the spin's magnetic field is either antiparallel (diamagnetic) or parallel (paramagnetic) to the spin's magnetic field.

Using Eq.~(\ref{eq:Emag}) the magnetic energy
\begin{align}
E_{\text{mag}} = \frac{16}{3} & \mu_0 \mu_r^{\text{I}} \mu_B^2 \pi \int_0^R {\cal P}(r)^2 r^2 dr \notag \\
&+ \left(\frac{\mu_r^{\text{II}}-\mu_r^{\text{I}}}{\mu_r^{\text{II}}+\mu_r^{\text{I}}}\right)\frac{\mu_0 \mu_r^{\text{I}} \mu_B^2 J^2}{32 \pi d^3} \left[3+\cos 2 \eta\right]. \label{eq:Emag3}
\end{align}
The first term is the magnetic energy of the spin itself, and is inversely proportional to $R^3$ (for ${\cal P}(r)=\left[j_0(\pi r/R)\right]^2$, a spherical Bessel function of zeroth order). The magnetic self-energy is experimentally inaccessible and goes to infinity for $R\rightarrow 0$, a well-known problem in classical electrodynamics~\cite{Jackson1998,Landau1971}. The second term in Eq.~(\ref{eq:Emag3}) represents the change to the magnetic energy due to the presence of region II. These corrections are independent of ${\cal P}(r)$ due to the assumed spherical symmetry. The other dependencies of the magnetic energy are trivial to understand, after realizing that the change in magnetic energy depends on how much of the spin's magnetic induction penetrates region II. The magnitude of the angular variation of the magnetic energy for $d=1$~nm is of the order of 10~neV (or 0.2~mK), which is extremely challenging to measure by spectroscopy. Also, the resulting force ${\bf F} = - \nabla E_{\text{mag}} \approx$~aN exerted on the scanning probe would be difficult to detect by atomic force microscopy. Instead, we will show that the magnetic energy can be probed using a coherent measurement of an NV$^-$ center's spin.

The ground state of an NV's spin $J=1$ is effectively described using the Hamiltonian ${\cal H}_{\text{NV}} = D_{\text{GS}} J_z^2$, where $D_{\text{GS}}\approx2.87$~GHz is the fine-structure constant due to the crystal field, and the $z$ direction is the NV$^-$ center's symmetry axis~\cite{Doherty2013}, see Fig.~\ref{fig:cartoon}(a). To compare the NV$^-$-center spin with the spin considered in Fig.~\ref{fig:Bfield}(a), it is convenient to orient the NV$^-$ center's symmetry axis perpendicular to the interface between the two regions. It has recently been demonstrated that such orientation can be realized deterministically in practice~\cite{Michl2014}. Analogous to the spin considered in Fig.~\ref{fig:Bfield}(a), the NV$^-$ center's spin is placed in the superposition $|J_\eta\rangle = \cos^2 \left(\eta/2\right) |+1\rangle + \tfrac{1}{2}\sqrt{2}\sin \eta |0\rangle + \sin^2 \left(\eta/2\right) |-1\rangle$, such that the expectation value of the spin makes an angle $\eta$ with respect to the $z$ axis. The energy of this state
\begin{align}
\langle J_\eta | {\cal H}_{\text{NV}} | J_\eta \rangle = \frac{D_{\text{GS}}}{4}\left[3+\cos2\eta\right]
\end{align}
is identical to the angular dependence of the magnetic energy in Eq.~(\ref{eq:Emag3}). Therefore the effect of a nearby region with different magnetic permeability on the spin of an NV$^-$ center seems to effectively change its fine-structure constant.

A fully quantum-mechanical treatment of the spin results in the magnetic energy Hamiltonian (see Supplemental Material~\cite{SM})
\begin{align}
{\cal H}_{\text{mag}} = \left(\frac{\mu_r^{\text{II}}-\mu_r^\text{I}}{\mu_r^{\text{II}}+\mu_r^\text{I}}\right) \frac{\mu_0 \mu_r^{\text{I}} \mu_B^2}{16 \pi \hbar^2 d^3} J_z^2 = D_{\text{mag}} J_z^2,
\end{align}
so that the NV$^-$ center effectively has $D=D_{\text{GS}}+D_{\text{mag}}$. Since the magnetization induced in region II depends on the spin and acts back on the spin itself, ${\cal H}_{\text{mag}}$ depends on the spin squared. In the Supplemental Material~\cite{SM} we show that ${\cal H}_{\text{mag}}$ has a similar structure when ${\cal P}({\bf x})$ has cylindrical symmetry and its axial symmetry axis is perpendicular to the interface between regions I and II. We also calculated that cylindrical symmetry changes an NV$^-$ center's $D_{\text{mag}}$ by $\leq 5\%$ from the spherical approximation. Lowering the symmetry further to NV$^-$'s $C_{3v}$ symmetry leads to additional small corrections, which we estimate to be less than $20\%$ for an NV$^-$ center 1~nm away from the interface. Assuming a spherical ${\cal P}({\bf x})$ is therefore a reasonable approximation. Note that in the classical limit $J \rightarrow \infty$ we get $\langle J_{\eta} | {\cal H}_{\text{mag}} | J_{\eta} \rangle = E_{\text{mag}}$, and also there is no effect for $J=\tfrac{1}{2}$.

The following (briefly outlined) coherent measurement protocol can be used to sensitively measure $D$; more details can be found in Ref.~\cite{Toyli2013}. The NV$^-$ center is first prepared in the $|J_z=0\rangle$ state using a pulsed optical excitation, by making use of the spin-dependent decay from the excited state manifold to the ground state manifold~\cite{Doherty2013}. The spin is then placed in a superposition of the $|J_z=0\rangle$ and $|J_z=\pm1\rangle$ states using a ${\pi}/{2}$ microwave pulse at frequency $D$. This superposition will acquire a phase $\exp({-iD\tau})$ after a free evolution time $\tau$. By applying another ${\pi}/{2}$ microwave pulse to project the spin onto the $|J_z=0\rangle$ state, the phase can be determined by optical measurement of the $|J_z=0\rangle$ population; $D$ follows from measuring the phase as function of $\tau$, most accurately through the use of a reference oscillator. The spin will experience decoherence during its free evolution; this can be  mitigated using dynamic decoupling protocols, which can be designed to optimize the sensitivity at which $D$ can be measured~\cite{Toyli2013}.

\begin{figure}
\includegraphics[width=\columnwidth]{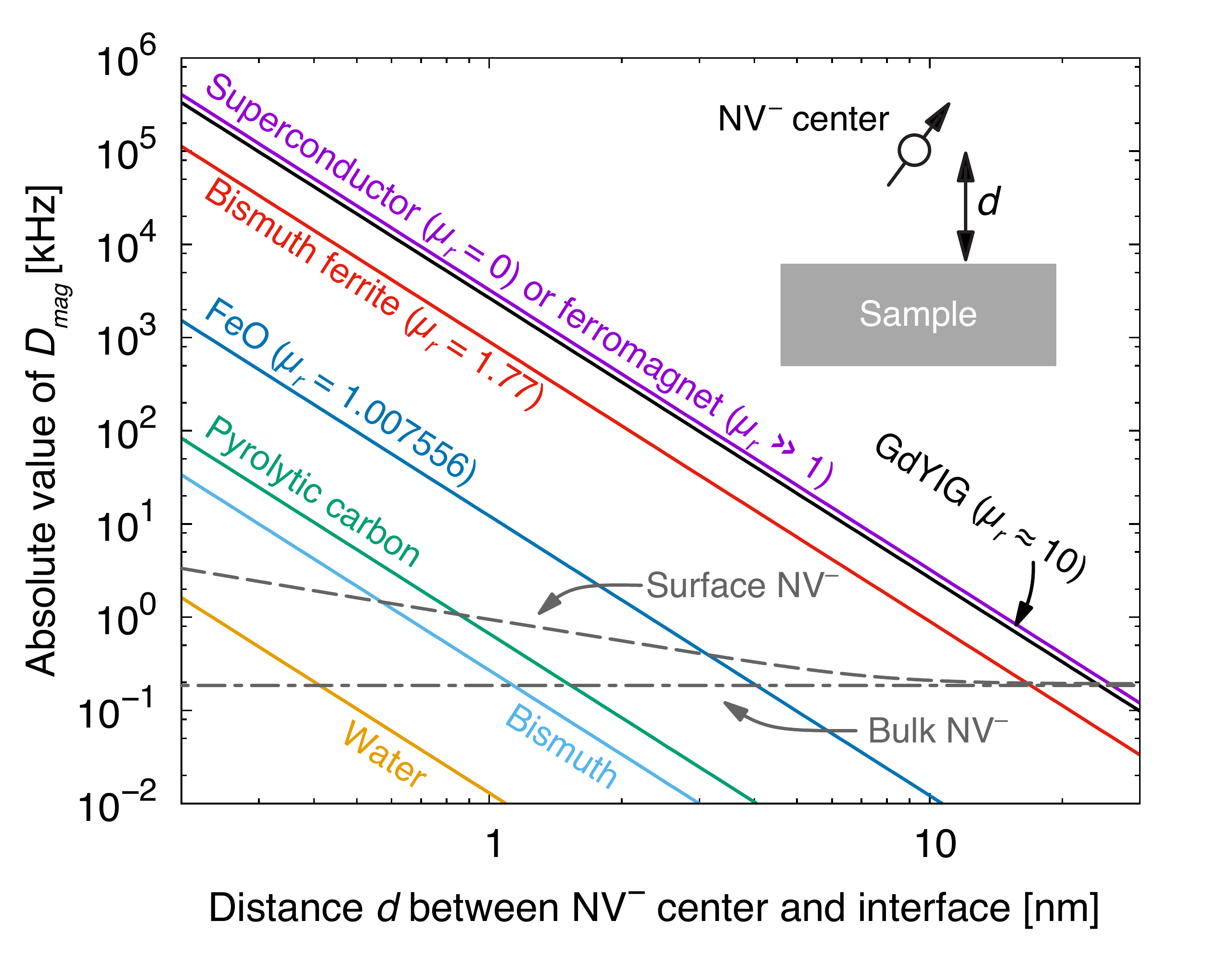}
\caption{The magnetic energy contribution $D_{\text{mag}}$ to the fine-structure constant as function of the distance $d$ between the NV$^-$ center and the sample, for samples having different magnetic permeabilities. The grey lines indicate the measurable change in $D$ for a measurement time of $100$s, reported for a bulk NV$^-$ center (dot-dash) and estimated for surface NV$^-$ centers (dash); see text for details. We took the low-frequency values for $\mu_r$ and assumed the superconductor to be a perfect diamagnet (i.e., vanishing penetration depth). The $\mu_r$ of pyrolytic carbon, bismuth, and water are, respectively, 0.999590, 0.999834, and 0.999992.}
\label{fig:detect}
\end{figure}

In Fig.~\ref{fig:detect} we show how $D_{\text{mag}}$ depends both on the distance $d$ between the NV$^-$ center and the sample, and on the relative magnetic permeability of the sample. Diamond itself has a very weak diamagnetic response, $\mu_r=1-2.2\times10^{-5}$~\cite{HBCP}, and has no free carriers. The NV$^-$ center is therefore in practice magnetically insensitive to its host, and $D_{\text{mag}}$ is barely affected by the diamond's shape. Using a coherent measurement technique, $D$ has been measured with a sensitivity of $1.85~\text{kHz}/\sqrt{\text{Hz}}$~\cite{Toyli2013}. Assuming a measurement time of 100 s, changes in $D$ of 0.2~kHz can therefore be detected for a bulk NV$^-$ center. From Fig.~\ref{fig:detect} it appears possible to detect both paramagnetic and diamagnetic substances if the NV$^-$ center is a few nm away from the sample. Such small distances are conventional in scanning probe microscopy~\cite{Yongho2008}, and have been achieved in conventional NV$^-$-center magnetometry~\cite{Loretz2014}. Recent studies showed that the proximity of the surface lowers the NV$^-$ center's $T_2$ coherence time due to a surface electronic spin bath and/or a surface phonon-related mechanism~\cite{Rosskopf2014,Myers2014,Romach2015}. This increases the minimal detectable change in $D$ by a factor $\sqrt{T_2^{\text{bulk}}/T_2^{\text{surface}}}$~\cite{Toyli2013}. Based on the experimental data of Ref.~\cite{Romach2015}, we roughly estimated the dependence of this ratio on $d$. We included in Fig.~\ref{fig:detect} both the minimal detectable change in $D$ estimated for near-surface and reported for bulk NV$^-$ centers for a measurement time of $100$ s. Increasing $T_2$ (potentially by mitigating the surface phenomena by surface passivation), improving sensing schemes, or extending the measurement time would push the minimal detectable change in $D$ down. 

\begin{figure}
\includegraphics[width=\columnwidth]{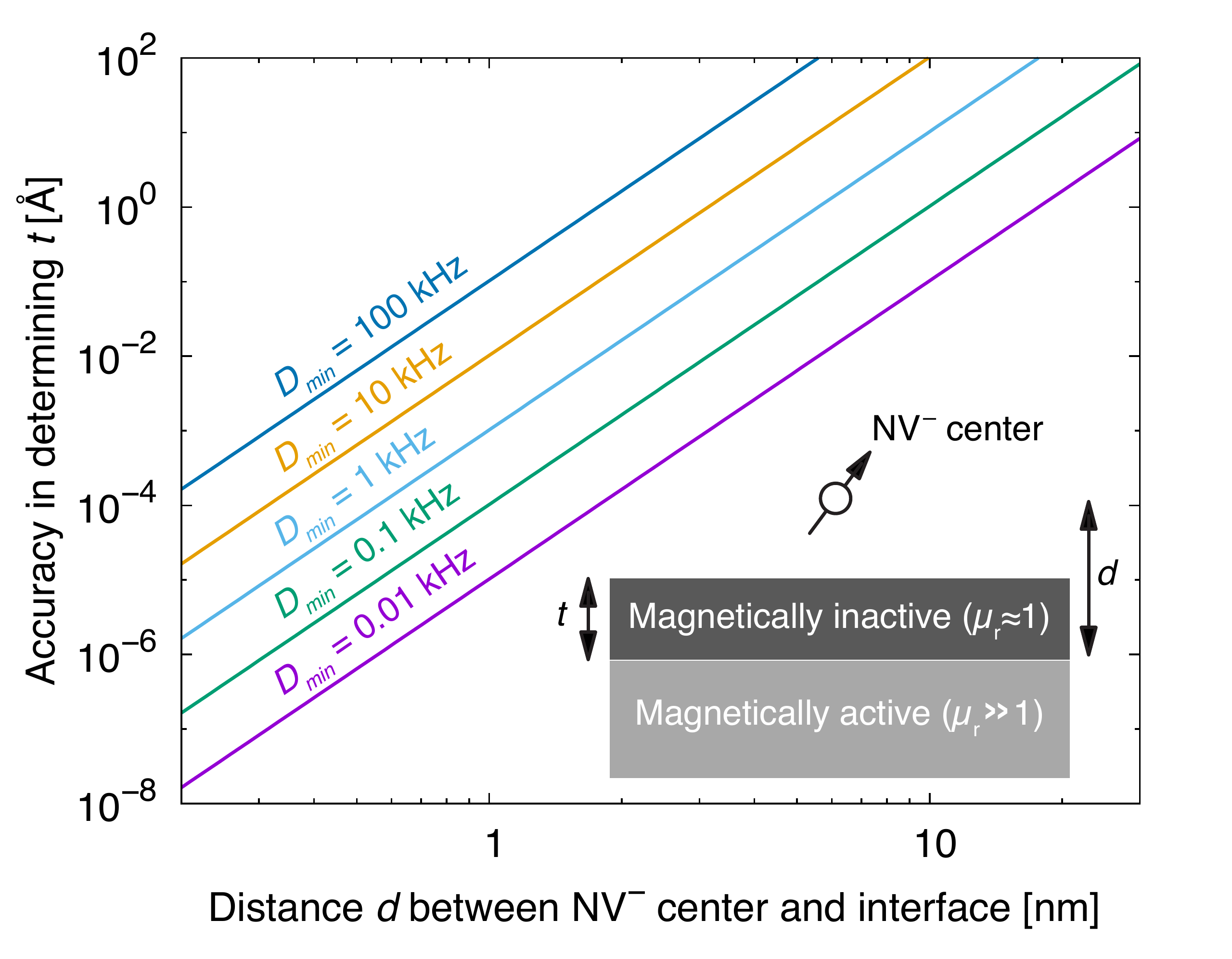}
\caption{The accuracy at which the thickness $t$ of a magnetically dead layer ($\mu_r$ is about 1) can be determined, as a function of the distance $d$ between the NV$^-$-center and the magnetically active region ($\mu_r\gg1$), for different minimal detectable changes $D_{\text{min}}$. Based on Ref.~\onlinecite{Toyli2013}, $D_{\text{min}} = 0.2$~kHz for a bulk NV$^-$-center and a measurement time of 100s. See Fig.~\ref{fig:detect} and its discussion in the text to determine an estimated $D_{\text{min}}$ close to the surface.}
\label{fig:sense}
\end{figure}

Our analysis is not limited to NV$^-$ centers; any spin close to a region with different magnetic permeability will experience an orientation-dependent magnetic energy, which affects its dynamics. Therefore the spins of other promising color centers~\cite{Gordon2013}, notably the divacancy in SiC, could also be used to detect the magnetic properties of nearby materials. Such systems would preferably have a smaller fine-structure constant $D_{\text{GS}}$, since in the proposed measurement scheme the NV$^-$ center's spin is precessing at that frequency. Although this does not impede the effect of the magnetic energy on the fine-structure constant, it does set the frequency at which the magnetic properties of the sample are probed; lowering this frequency would be favorable. Alternatively, different measurement schemes could be developed, which remove the necessity of the spin precessing at such frequencies.

As an example of the added value of the proposed magnetic-energy-based magnetometry, we suggest to use this technique to measure the thickness $t$ of magnetically dead layers~\cite{Sun1999}. A common problem in magnetic multilayered materials, such as magnetic tunnel junctions~\cite{Ikeda2010}, is the magnetic inactivity of the top surface layer of the structure, see inset of Fig.~\ref{fig:sense}. We can make use of the strong distance dependence of the magnetic energy ($E_{\text{mag}} \propto d^{-3}$) to sensitively determine the distance $d$ between the NV$^-$ center and the boundary of the magnetically active material. As the separation between the NV$^-$ center and the physical boundary of the sample is known through calibration, the thickness $t$ of the magnetically inactive material can be determined with high precision. Figure~\ref{fig:sense} predicts that this can be achieved with remarkable accuracy.

We  propose a method to sense the magnetic properties of materials based on the magnetic energy of a nearby spin. This method inverts the conventional scheme of scanning-probe magnetometers, making it possible to sense materials which have no natural magnetic field external to their volume. This scheme can be applied to NV$^-$ centers and, using realistic assumptions, we predict it should be possible to detect both para- and diamagnetic materials. Future theoretical work towards implementing different color centers or different measurement schemes could lower the frequency at which the magnetic properties are probed and improve the predicted sensitivity.  

The authors acknowledge support from an Air Force Office of Scientific Research (AFOSR) Multidisciplinary University Research Initiative (MURI) Grant and J. v. B. acknowledges a Rubicon Grant from the Netherlands Organization for Scientific Research.


\begin{thebibliography}{33}%
\makeatletter
\providecommand \@ifxundefined [1]{%
 \@ifx{#1\undefined}
}%
\providecommand \@ifnum [1]{%
 \ifnum #1\expandafter \@firstoftwo
 \else \expandafter \@secondoftwo
 \fi
}%
\providecommand \@ifx [1]{%
 \ifx #1\expandafter \@firstoftwo
 \else \expandafter \@secondoftwo
 \fi
}%
\providecommand \natexlab [1]{#1}%
\providecommand \enquote  [1]{``#1''}%
\providecommand \bibnamefont  [1]{#1}%
\providecommand \bibfnamefont [1]{#1}%
\providecommand \citenamefont [1]{#1}%
\providecommand \href@noop [0]{\@secondoftwo}%
\providecommand \href [0]{\begingroup \@sanitize@url \@href}%
\providecommand \@href[1]{\@@startlink{#1}\@@href}%
\providecommand \@@href[1]{\endgroup#1\@@endlink}%
\providecommand \@sanitize@url [0]{\catcode `\\12\catcode `\$12\catcode
  `\&12\catcode `\#12\catcode `\^12\catcode `\_12\catcode `\%12\relax}%
\providecommand \@@startlink[1]{}%
\providecommand \@@endlink[0]{}%
\providecommand \url  [0]{\begingroup\@sanitize@url \@url }%
\providecommand \@url [1]{\endgroup\@href {#1}{\urlprefix }}%
\providecommand \urlprefix  [0]{URL }%
\providecommand \Eprint [0]{\href }%
\providecommand \doibase [0]{http://dx.doi.org/}%
\providecommand \selectlanguage [0]{\@gobble}%
\providecommand \bibinfo  [0]{\@secondoftwo}%
\providecommand \bibfield  [0]{\@secondoftwo}%
\providecommand \translation [1]{[#1]}%
\providecommand \BibitemOpen [0]{}%
\providecommand \bibitemStop [0]{}%
\providecommand \bibitemNoStop [0]{.\EOS\space}%
\providecommand \EOS [0]{\spacefactor3000\relax}%
\providecommand \BibitemShut  [1]{\csname bibitem#1\endcsname}%
\let\auto@bib@innerbib\@empty
\bibitem [{\citenamefont {Rabi}\ \emph {et~al.}(1938)\citenamefont {Rabi},
  \citenamefont {Zacharias}, \citenamefont {Millman},\ and\ \citenamefont
  {Kusch}}]{Rabi1938}%
  \BibitemOpen
  \bibfield  {author} {\bibinfo {author} {\bibfnamefont {I.~I.}\ \bibnamefont
  {Rabi}}, \bibinfo {author} {\bibfnamefont {J.~R.}\ \bibnamefont {Zacharias}},
  \bibinfo {author} {\bibfnamefont {S.}~\bibnamefont {Millman}}, \ and\
  \bibinfo {author} {\bibfnamefont {P.}~\bibnamefont {Kusch}},\ }\href@noop {}
  {\bibfield  {journal} {\bibinfo  {journal} {Phys. Rev.}\ }\textbf {\bibinfo
  {volume} {53}},\ \bibinfo {pages} {318} (\bibinfo {year} {1938})}\BibitemShut
  {NoStop}%
\bibitem [{\citenamefont {Jaklevic}\ \emph {et~al.}(1964)\citenamefont
  {Jaklevic}, \citenamefont {Lambe}, \citenamefont {Silver},\ and\
  \citenamefont {Mercereau}}]{Jaklevic1964}%
  \BibitemOpen
  \bibfield  {author} {\bibinfo {author} {\bibfnamefont {R.~C.}\ \bibnamefont
  {Jaklevic}}, \bibinfo {author} {\bibfnamefont {J.}~\bibnamefont {Lambe}},
  \bibinfo {author} {\bibfnamefont {A.~H.}\ \bibnamefont {Silver}}, \ and\
  \bibinfo {author} {\bibfnamefont {J.~E.}\ \bibnamefont {Mercereau}},\
  }\href@noop {} {\bibfield  {journal} {\bibinfo  {journal} {Phys. Rev. Lett.}\
  }\textbf {\bibinfo {volume} {12}},\ \bibinfo {pages} {159} (\bibinfo {year}
  {1964})}\BibitemShut {NoStop}%
\bibitem [{\citenamefont {Wehrli}(1992)}]{Wehrli1992}%
  \BibitemOpen
  \bibfield  {author} {\bibinfo {author} {\bibfnamefont {F.~W.}\ \bibnamefont
  {Wehrli}},\ }\href@noop {} {\bibfield  {journal} {\bibinfo  {journal}
  {Phys. Today}\ }\textbf {\bibinfo {volume} {45}},\ \bibinfo {pages} {34}
  (\bibinfo {year} {1992})}\BibitemShut {NoStop}%
\bibitem [{\citenamefont {Balasubramanian}\ \emph {et~al.}(2008)\citenamefont
  {Balasubramanian}, \citenamefont {Chan}, \citenamefont {Kolesov},
  \citenamefont {Al-Hmoud}, \citenamefont {Tisler}, \citenamefont {Shin},
  \citenamefont {Kim}, \citenamefont {Wojcik}, \citenamefont {Hemmer},
  \citenamefont {Krueger}, \citenamefont {Hanke}, \citenamefont
  {Leitenstorfer}, \citenamefont {Bratschitsch}, \citenamefont {Jelezko},\ and\
  \citenamefont {Wrachtrup}}]{Balasubramanian2008}%
  \BibitemOpen
  \bibfield  {author} {\bibinfo {author} {\bibfnamefont {G.}~\bibnamefont
  {Balasubramanian}}, \bibinfo {author} {\bibfnamefont {I.~Y.}\ \bibnamefont
  {Chan}}, \bibinfo {author} {\bibfnamefont {R.}~\bibnamefont {Kolesov}},
  \bibinfo {author} {\bibfnamefont {M.}~\bibnamefont {Al-Hmoud}}, \bibinfo
  {author} {\bibfnamefont {J.}~\bibnamefont {Tisler}}, \bibinfo {author}
  {\bibfnamefont {C.}~\bibnamefont {Shin}}, \bibinfo {author} {\bibfnamefont
  {C.}~\bibnamefont {Kim}}, \bibinfo {author} {\bibfnamefont {A.}~\bibnamefont
  {Wojcik}}, \bibinfo {author} {\bibfnamefont {P.~R.}\ \bibnamefont {Hemmer}},
  \bibinfo {author} {\bibfnamefont {A.}~\bibnamefont {Krueger}}, \bibinfo
  {author} {\bibfnamefont {T.}~\bibnamefont {Hanke}}, \bibinfo {author}
  {\bibfnamefont {A.}~\bibnamefont {Leitenstorfer}}, \bibinfo {author}
  {\bibfnamefont {R.}~\bibnamefont {Bratschitsch}}, \bibinfo {author}
  {\bibfnamefont {F.}~\bibnamefont {Jelezko}}, \ and\ \bibinfo {author}
  {\bibfnamefont {J.}~\bibnamefont {Wrachtrup}},\ }\href@noop {} {\bibfield
  {journal} {\bibinfo  {journal} {Nature (London)}\ }\textbf {\bibinfo {volume} {455}},\
  \bibinfo {pages} {648} (\bibinfo {year} {2008})}\BibitemShut {NoStop}%
\bibitem [{\citenamefont {Ciobanu}\ \emph {et~al.}(2002)\citenamefont
  {Ciobanu}, \citenamefont {Seeber},\ and\ \citenamefont
  {Pennington}}]{Ciobanu2002}%
  \BibitemOpen
  \bibfield  {author} {\bibinfo {author} {\bibfnamefont {L.}~\bibnamefont
  {Ciobanu}}, \bibinfo {author} {\bibfnamefont {D.}~\bibnamefont {Seeber}}, \
  and\ \bibinfo {author} {\bibfnamefont {C.}~\bibnamefont {Pennington}},\
  }\href@noop {} {\bibfield  {journal} {\bibinfo  {journal} {J. Magn. Reson.}\ }\textbf {\bibinfo {volume} {158}},\ \bibinfo {pages}
  {178} (\bibinfo {year} {2002})}\BibitemShut {NoStop}%
\bibitem [{\citenamefont {Glover}\ and\ \citenamefont
  {Mansfield}(2002)}]{Glover2002}%
  \BibitemOpen
  \bibfield  {author} {\bibinfo {author} {\bibfnamefont {P.}~\bibnamefont
  {Glover}}\ and\ \bibinfo {author} {\bibfnamefont {S.~P.}\ \bibnamefont
  {Mansfield}},\ }\href@noop {} {\bibfield  {journal} {\bibinfo  {journal}
  {Rep. Prog. Phys.}\ }\textbf {\bibinfo {volume} {65}},\
  \bibinfo {pages} {1489} (\bibinfo {year} {2002})}\BibitemShut {NoStop}%
\bibitem [{\citenamefont {Martin}\ and\ \citenamefont
  {Wickramasinghe}(1987)}]{Martin1987}%
  \BibitemOpen
  \bibfield  {author} {\bibinfo {author} {\bibfnamefont {Y.}~\bibnamefont
  {Martin}}\ and\ \bibinfo {author} {\bibfnamefont {H.~K.}\ \bibnamefont
  {Wickramasinghe}},\ }\href@noop {} {\bibfield  {journal} {\bibinfo  {journal}
  {Appl. Phys. Lett.}\ }\textbf {\bibinfo {volume} {50}},\ \bibinfo
  {pages} {1455} (\bibinfo {year} {1987})}\BibitemShut {NoStop}%
\bibitem [{\citenamefont {Chang}\ \emph {et~al.}(1992)\citenamefont {Chang},
  \citenamefont {Hallen}, \citenamefont {Harriott}, \citenamefont {Hess},
  \citenamefont {Kao}, \citenamefont {Kwo}, \citenamefont {Miller},
  \citenamefont {Wolfe}, \citenamefont {van~der Ziel},\ and\ \citenamefont
  {Chang}}]{Chang1992}%
  \BibitemOpen
  \bibfield  {author} {\bibinfo {author} {\bibfnamefont {A.~M.}\ \bibnamefont
  {Chang}}, \bibinfo {author} {\bibfnamefont {H.~D.}\ \bibnamefont {Hallen}},
  \bibinfo {author} {\bibfnamefont {L.}~\bibnamefont {Harriott}}, \bibinfo
  {author} {\bibfnamefont {H.~F.}\ \bibnamefont {Hess}}, \bibinfo {author}
  {\bibfnamefont {H.~L.}\ \bibnamefont {Kao}}, \bibinfo {author} {\bibfnamefont
  {J.}~\bibnamefont {Kwo}}, \bibinfo {author} {\bibfnamefont {R.~E.}\
  \bibnamefont {Miller}}, \bibinfo {author} {\bibfnamefont {R.}~\bibnamefont
  {Wolfe}}, \bibinfo {author} {\bibfnamefont {J.}~\bibnamefont {van~der Ziel}},
  \ and\ \bibinfo {author} {\bibfnamefont {T.~Y.}\ \bibnamefont {Chang}},\
  }\href@noop {} {\bibfield  {journal} {\bibinfo  {journal} {Appl. Phys.
  Lett.}\ }\textbf {\bibinfo {volume} {61}},\ \bibinfo {pages} {1974}
  (\bibinfo {year} {1992})}\BibitemShut {NoStop}%
\bibitem [{\citenamefont {Degen}\ \emph {et~al.}(2009)\citenamefont {Degen},
  \citenamefont {Poggio}, \citenamefont {Mamin}, \citenamefont {Rettner},\ and\
  \citenamefont {Rugar}}]{Degen2009}%
  \BibitemOpen
  \bibfield  {author} {\bibinfo {author} {\bibfnamefont {C.~L.}\ \bibnamefont
  {Degen}}, \bibinfo {author} {\bibfnamefont {M.}~\bibnamefont {Poggio}},
  \bibinfo {author} {\bibfnamefont {H.~J.}\ \bibnamefont {Mamin}}, \bibinfo
  {author} {\bibfnamefont {C.~T.}\ \bibnamefont {Rettner}}, \ and\ \bibinfo
  {author} {\bibfnamefont {D.}~\bibnamefont {Rugar}},\ }\href@noop {}
  {\bibfield  {journal} {\bibinfo  {journal} {Proc. Natl.
  Acad. Sci. U.S.A.}\ }\textbf {\bibinfo {volume} {106}},\ \bibinfo {pages}
  {1313} (\bibinfo {year} {2009})}\BibitemShut {NoStop}%
\bibitem [{\citenamefont {Hong}\ \emph {et~al.}(2013)\citenamefont {Hong},
  \citenamefont {Grinolds}, \citenamefont {Pham}, \citenamefont {Le~Sage},
  \citenamefont {Luan}, \citenamefont {Walsworth},\ and\ \citenamefont
  {Yacoby}}]{Hong2013}%
  \BibitemOpen
  \bibfield  {author} {\bibinfo {author} {\bibfnamefont {S.}~\bibnamefont
  {Hong}}, \bibinfo {author} {\bibfnamefont {M.~S.}\ \bibnamefont {Grinolds}},
  \bibinfo {author} {\bibfnamefont {L.~M.}\ \bibnamefont {Pham}}, \bibinfo
  {author} {\bibfnamefont {D.}~\bibnamefont {Le~Sage}}, \bibinfo {author}
  {\bibfnamefont {L.}~\bibnamefont {Luan}}, \bibinfo {author} {\bibfnamefont
  {R.~L.}\ \bibnamefont {Walsworth}}, \ and\ \bibinfo {author} {\bibfnamefont
  {A.}~\bibnamefont {Yacoby}},\ }\href@noop {} {\bibfield  {journal} {\bibinfo
  {journal} {MRS Bull.}\ }\textbf {\bibinfo {volume} {38}},\ \bibinfo
  {pages} {155} (\bibinfo {year} {2013})}\BibitemShut {NoStop}%
\bibitem [{\citenamefont {van Oort}\ \emph {et~al.}(1988)\citenamefont {van
  Oort}, \citenamefont {Manson},\ and\ \citenamefont {Glasbeek}}]{Oort1988}%
  \BibitemOpen
  \bibfield  {author} {\bibinfo {author} {\bibfnamefont {E.}~\bibnamefont {van
  Oort}}, \bibinfo {author} {\bibfnamefont {N.~B.}\ \bibnamefont {Manson}}, \
  and\ \bibinfo {author} {\bibfnamefont {M.}~\bibnamefont {Glasbeek}},\
  }\href@noop {} {\bibfield  {journal} {\bibinfo  {journal} {J. Phys.
  C}\ }\textbf {\bibinfo {volume} {21}},\ \bibinfo {pages}
  {4385} (\bibinfo {year} {1988})}\BibitemShut {NoStop}%
\bibitem [{\citenamefont {Childress}\ \emph {et~al.}(2006)\citenamefont
  {Childress}, \citenamefont {Dutt}, \citenamefont {Taylor}, \citenamefont
  {Zibrov}, \citenamefont {Jelezko}, \citenamefont {Wrachtrup}, \citenamefont
  {Hemmer},\ and\ \citenamefont {Lukin}}]{Childress2006}%
  \BibitemOpen
  \bibfield  {author} {\bibinfo {author} {\bibfnamefont {L.}~\bibnamefont
  {Childress}}, \bibinfo {author} {\bibfnamefont {M.~V.~G.}\ \bibnamefont
  {Dutt}}, \bibinfo {author} {\bibfnamefont {J.~M.}\ \bibnamefont {Taylor}},
  \bibinfo {author} {\bibfnamefont {A.~S.}\ \bibnamefont {Zibrov}}, \bibinfo
  {author} {\bibfnamefont {F.}~\bibnamefont {Jelezko}}, \bibinfo {author}
  {\bibfnamefont {J.}~\bibnamefont {Wrachtrup}}, \bibinfo {author}
  {\bibfnamefont {P.~R.}\ \bibnamefont {Hemmer}}, \ and\ \bibinfo {author}
  {\bibfnamefont {M.~D.}\ \bibnamefont {Lukin}},\ }\href {\doibase
  10.1126/science.1131871} {\bibfield  {journal} {\bibinfo  {journal}
  {Science}\ }\textbf {\bibinfo {volume} {314}},\ \bibinfo {pages} {281}
  (\bibinfo {year} {2006})}\BibitemShut {NoStop}%
\bibitem [{\citenamefont {Maze}\ \emph {et~al.}(2008)\citenamefont {Maze},
  \citenamefont {Stanwix}, \citenamefont {Hodges}, \citenamefont {Hong},
  \citenamefont {Taylor}, \citenamefont {Cappellaro}, \citenamefont {Jiang},
  \citenamefont {Dutt}, \citenamefont {Togan}, \citenamefont {Zibrov},
  \citenamefont {Yacoby}, \citenamefont {Walsworth},\ and\ \citenamefont
  {Lukin}}]{Maze2008}%
  \BibitemOpen
  \bibfield  {author} {\bibinfo {author} {\bibfnamefont {J.~R.}\ \bibnamefont
  {Maze}}, \bibinfo {author} {\bibfnamefont {P.~L.}\ \bibnamefont {Stanwix}},
  \bibinfo {author} {\bibfnamefont {J.~S.}\ \bibnamefont {Hodges}}, \bibinfo
  {author} {\bibfnamefont {S.}~\bibnamefont {Hong}}, \bibinfo {author}
  {\bibfnamefont {J.~M.}\ \bibnamefont {Taylor}}, \bibinfo {author}
  {\bibfnamefont {P.}~\bibnamefont {Cappellaro}}, \bibinfo {author}
  {\bibfnamefont {L.}~\bibnamefont {Jiang}}, \bibinfo {author} {\bibfnamefont
  {M.~V.~G.}\ \bibnamefont {Dutt}}, \bibinfo {author} {\bibfnamefont
  {E.}~\bibnamefont {Togan}}, \bibinfo {author} {\bibfnamefont {A.~S.}\
  \bibnamefont {Zibrov}}, \bibinfo {author} {\bibfnamefont {A.}~\bibnamefont
  {Yacoby}}, \bibinfo {author} {\bibfnamefont {R.~L.}\ \bibnamefont
  {Walsworth}}, \ and\ \bibinfo {author} {\bibfnamefont {M.~D.}\ \bibnamefont
  {Lukin}},\ }\href@noop {} {\bibfield  {journal} {\bibinfo  {journal}
  {Nature (London)}\ }\textbf {\bibinfo {volume} {455}},\ \bibinfo {pages} {644}
  (\bibinfo {year} {2008})}\BibitemShut {NoStop}%
\bibitem [{\citenamefont {Toyli}\ \emph {et~al.}(2013)\citenamefont {Toyli},
  \citenamefont {de~las Casas}, \citenamefont {Christle}, \citenamefont
  {Dobrovitski},\ and\ \citenamefont {Awschalom}}]{Toyli2013}%
  \BibitemOpen
  \bibfield  {author} {\bibinfo {author} {\bibfnamefont {D.~M.}\ \bibnamefont
  {Toyli}}, \bibinfo {author} {\bibfnamefont {C.~F.}\ \bibnamefont {de~las
  Casas}}, \bibinfo {author} {\bibfnamefont {D.~J.}\ \bibnamefont {Christle}},
  \bibinfo {author} {\bibfnamefont {V.~V.}\ \bibnamefont {Dobrovitski}}, \ and\
  \bibinfo {author} {\bibfnamefont {D.~D.}\ \bibnamefont {Awschalom}},\
  }\href@noop {} {\bibfield  {journal} {\bibinfo  {journal} {Proc. Natl. Acad. Sci. U.S.A.}\ }\textbf {\bibinfo {volume} {110}},\ \bibinfo
  {pages} {8417} (\bibinfo {year} {2013})}\BibitemShut {NoStop}%
\bibitem [{\citenamefont {Kucsko}\ \emph {et~al.}(2013)\citenamefont {Kucsko},
  \citenamefont {Maurer}, \citenamefont {Yao}, \citenamefont {Kubo},
  \citenamefont {Noh}, \citenamefont {Lo}, \citenamefont {Park},\ and\
  \citenamefont {Lukin}}]{Kucsko2013}%
  \BibitemOpen
  \bibfield  {author} {\bibinfo {author} {\bibfnamefont {G.}~\bibnamefont
  {Kucsko}}, \bibinfo {author} {\bibfnamefont {P.~C.}\ \bibnamefont {Maurer}},
  \bibinfo {author} {\bibfnamefont {N.~Y.}\ \bibnamefont {Yao}}, \bibinfo
  {author} {\bibfnamefont {M.}~\bibnamefont {Kubo}}, \bibinfo {author}
  {\bibfnamefont {H.~J.}\ \bibnamefont {Noh}}, \bibinfo {author} {\bibfnamefont
  {P.~K.}\ \bibnamefont {Lo}}, \bibinfo {author} {\bibfnamefont
  {H.}~\bibnamefont {Park}}, \ and\ \bibinfo {author} {\bibfnamefont {M.~D.}\
  \bibnamefont {Lukin}},\ }\href@noop {} {\bibfield  {journal} {\bibinfo
  {journal} {Nature (London)}\ }\textbf {\bibinfo {volume} {500}},\ \bibinfo {pages}
  {54} (\bibinfo {year} {2013})}\BibitemShut {NoStop}%
\bibitem [{\citenamefont {Neumann}\ \emph {et~al.}(2013)\citenamefont
  {Neumann}, \citenamefont {Jakobi}, \citenamefont {Dolde}, \citenamefont
  {Burk}, \citenamefont {Reuter}, \citenamefont {Waldherr}, \citenamefont
  {Honert}, \citenamefont {Wolf}, \citenamefont {Brunner}, \citenamefont
  {Shim}, \citenamefont {Suter}, \citenamefont {Sumiya}, \citenamefont
  {Isoya},\ and\ \citenamefont {Wrachtrup}}]{Neumann2013}%
  \BibitemOpen
  \bibfield  {author} {\bibinfo {author} {\bibfnamefont {P.}~\bibnamefont
  {Neumann}}, \bibinfo {author} {\bibfnamefont {I.}~\bibnamefont {Jakobi}},
  \bibinfo {author} {\bibfnamefont {F.}~\bibnamefont {Dolde}}, \bibinfo
  {author} {\bibfnamefont {C.}~\bibnamefont {Burk}}, \bibinfo {author}
  {\bibfnamefont {R.}~\bibnamefont {Reuter}}, \bibinfo {author} {\bibfnamefont
  {G.}~\bibnamefont {Waldherr}}, \bibinfo {author} {\bibfnamefont
  {J.}~\bibnamefont {Honert}}, \bibinfo {author} {\bibfnamefont
  {T.}~\bibnamefont {Wolf}}, \bibinfo {author} {\bibfnamefont {A.}~\bibnamefont
  {Brunner}}, \bibinfo {author} {\bibfnamefont {J.~H.}\ \bibnamefont {Shim}},
  \bibinfo {author} {\bibfnamefont {D.}~\bibnamefont {Suter}}, \bibinfo
  {author} {\bibfnamefont {H.}~\bibnamefont {Sumiya}}, \bibinfo {author}
  {\bibfnamefont {J.}~\bibnamefont {Isoya}}, \ and\ \bibinfo {author}
  {\bibfnamefont {J.}~\bibnamefont {Wrachtrup}},\ }\href@noop {} {\bibfield
  {journal} {\bibinfo  {journal} {Nano Lett.}\ }\textbf {\bibinfo {volume}
  {13}},\ \bibinfo {pages} {2738} (\bibinfo {year} {2013})}\BibitemShut
  {NoStop}%
\bibitem [{\citenamefont {Sun}\ \emph {et~al.}(1999)\citenamefont {Sun},
  \citenamefont {Abraham}, \citenamefont {Rao},\ and\ \citenamefont
  {Eom}}]{Sun1999}%
  \BibitemOpen
  \bibfield  {author} {\bibinfo {author} {\bibfnamefont {J.~Z.}\ \bibnamefont
  {Sun}}, \bibinfo {author} {\bibfnamefont {D.~W.}\ \bibnamefont {Abraham}},
  \bibinfo {author} {\bibfnamefont {R.~A.}\ \bibnamefont {Rao}}, \ and\
  \bibinfo {author} {\bibfnamefont {C.~B.}\ \bibnamefont {Eom}},\ }\href@noop
  {} {\bibfield  {journal} {\bibinfo  {journal} {Appl. Phys. Lett.}\
  }\textbf {\bibinfo {volume} {74}},\ \bibinfo {pages} {3017} (\bibinfo {year}
  {1999})}\BibitemShut {NoStop}%
\bibitem [{SM()}]{SM}%
  \BibitemOpen
  \href@noop {} {}\bibinfo {note} {See Supplemental Material for the
  derivation of the spin density of an $N$-particle state [validating Eq.~(1)],
  and for the derivation of the magnetic energy Hamiltonian treating the spin
  fully quantum mechanically and assuming a spherically or cylindrically
  symmetric probability density. The Supplemental Material
  includes Refs.~\cite{Schwabl2004,Bjorken1965,Gali2008}.}\BibitemShut {Stop}%
\bibitem [{\citenamefont {Doherty}\ \emph {et~al.}(2013)\citenamefont
  {Doherty}, \citenamefont {Manson}, \citenamefont {Delaney}, \citenamefont
  {Jelezko}, \citenamefont {Wrachtrup},\ and\ \citenamefont
  {Hollenberg}}]{Doherty2013}%
  \BibitemOpen
  \bibfield  {author} {\bibinfo {author} {\bibfnamefont {M.~W.}\ \bibnamefont
  {Doherty}}, \bibinfo {author} {\bibfnamefont {N.~B.}\ \bibnamefont {Manson}},
  \bibinfo {author} {\bibfnamefont {P.}~\bibnamefont {Delaney}}, \bibinfo
  {author} {\bibfnamefont {F.}~\bibnamefont {Jelezko}}, \bibinfo {author}
  {\bibfnamefont {J.}~\bibnamefont {Wrachtrup}}, \ and\ \bibinfo {author}
  {\bibfnamefont {L.~C.~L.}\ \bibnamefont {Hollenberg}},\ }\href@noop {}
  {\bibfield  {journal} {\bibinfo  {journal} {Phys. Rep.}\ }\textbf
  {\bibinfo {volume} {528}},\ \bibinfo {pages} {1} (\bibinfo {year}
  {2013})}\BibitemShut {NoStop}%
\bibitem [{\citenamefont {Jackson}(1998)}]{Jackson1998}%
  \BibitemOpen
  \bibfield  {author} {\bibinfo {author} {\bibfnamefont {J.~D.}\ \bibnamefont
  {Jackson}},\ }\href@noop {} {\emph {\bibinfo {title} {Classical
  Electrodynamics}}},\ \bibinfo {edition} {3rd}\ ed.\ (\bibinfo  {publisher}
  {Wiley},\ \bibinfo {address} {New York},\ \bibinfo {year} {1998})\BibitemShut
  {NoStop}%
\bibitem [{\citenamefont {Landau}\ and\ \citenamefont
  {Lifshitz}(1971)}]{Landau1971}%
  \BibitemOpen
  \bibfield  {author} {\bibinfo {author} {\bibfnamefont {L.~D.}\ \bibnamefont
  {Landau}}\ and\ \bibinfo {author} {\bibfnamefont {E.~M.}\ \bibnamefont
  {Lifshitz}},\ }\href@noop {} {\emph {\bibinfo {title} {The Classical Theory
  of Fields}}},\ \bibinfo {edition} {3rd}\ ed.,\ \bibinfo {series} {Landau and
  Lifshitz Course of Theoretical Physics} - Vol.~\bibinfo {volume} {2}\
  (\bibinfo  {publisher} {Pergamon Press}, Oxford,\ \bibinfo {year} {1971})\BibitemShut
  {NoStop}%
\bibitem [{\citenamefont {Michl}\ \emph {et~al.}(2014)\citenamefont {Michl},
  \citenamefont {Teraji}, \citenamefont {Zaiser}, \citenamefont {Jakobi},
  \citenamefont {Waldherr}, \citenamefont {Dolde}, \citenamefont {Neumann},
  \citenamefont {Doherty}, \citenamefont {Manson}, \citenamefont {Isoya},\ and\
  \citenamefont {Wrachtrup}}]{Michl2014}%
  \BibitemOpen
  \bibfield  {author} {\bibinfo {author} {\bibfnamefont {J.}~\bibnamefont
  {Michl}}, \bibinfo {author} {\bibfnamefont {T.}~\bibnamefont {Teraji}},
  \bibinfo {author} {\bibfnamefont {S.}~\bibnamefont {Zaiser}}, \bibinfo
  {author} {\bibfnamefont {I.}~\bibnamefont {Jakobi}}, \bibinfo {author}
  {\bibfnamefont {G.}~\bibnamefont {Waldherr}}, \bibinfo {author}
  {\bibfnamefont {F.}~\bibnamefont {Dolde}}, \bibinfo {author} {\bibfnamefont
  {P.}~\bibnamefont {Neumann}}, \bibinfo {author} {\bibfnamefont {M.~W.}\
  \bibnamefont {Doherty}}, \bibinfo {author} {\bibfnamefont {N.~B.}\
  \bibnamefont {Manson}}, \bibinfo {author} {\bibfnamefont {J.}~\bibnamefont
  {Isoya}}, \ and\ \bibinfo {author} {\bibfnamefont {J.}~\bibnamefont
  {Wrachtrup}},\ }\href@noop {} {\bibfield  {journal} {\bibinfo  {journal}
  {Appl. Phys. Lett.}\ }\textbf {\bibinfo {volume} {104}},\ \bibinfo
  {pages} {102407} (\bibinfo {year} {2014})}\BibitemShut {NoStop}%
\bibitem [{\citenamefont {Haynes}(2015)}]{HBCP}%
  \BibitemOpen
  \bibfield  {author} {\bibinfo {author} {\bibfnamefont {W.~M.}\ \bibnamefont
  {Haynes}},\ }\href@noop {} {\emph {\bibinfo {title} {Handbook of Chemistry
  and Physics}}}\ (\bibinfo  {publisher} {CRC}, Cleveland, \ \bibinfo {year}
  {2015})\BibitemShut {NoStop}%
\bibitem [{\citenamefont {Seo}\ and\ \citenamefont {Jhe}(2008)}]{Yongho2008}%
  \BibitemOpen
  \bibfield  {author} {\bibinfo {author} {\bibfnamefont {Y.}~\bibnamefont
  {Seo}}\ and\ \bibinfo {author} {\bibfnamefont {W.}~\bibnamefont {Jhe}},\
  }\href@noop {} {\bibfield  {journal} {\bibinfo  {journal} {Rep. Prog. Phys.}\ }\textbf {\bibinfo {volume} {71}},\ \bibinfo {pages}
  {016101} (\bibinfo {year} {2008})}\BibitemShut {NoStop}%
\bibitem [{\citenamefont {Loretz}\ \emph {et~al.}(2014)\citenamefont {Loretz},
  \citenamefont {Pezzagna}, \citenamefont {Meijer},\ and\ \citenamefont
  {Degen}}]{Loretz2014}%
  \BibitemOpen
  \bibfield  {author} {\bibinfo {author} {\bibfnamefont {M.}~\bibnamefont
  {Loretz}}, \bibinfo {author} {\bibfnamefont {S.}~\bibnamefont {Pezzagna}},
  \bibinfo {author} {\bibfnamefont {J.}~\bibnamefont {Meijer}}, \ and\ \bibinfo
  {author} {\bibfnamefont {C.~L.}\ \bibnamefont {Degen}},\ }\href@noop {}
  {\bibfield  {journal} {\bibinfo  {journal} {Appl. Phys. Lett.}\
  }\textbf {\bibinfo {volume} {104}},\ \bibinfo {pages} {033102} (\bibinfo
  {year} {2014})}\BibitemShut {NoStop}%
\bibitem [{\citenamefont {Rosskopf}\ \emph {et~al.}(2014)\citenamefont
  {Rosskopf}, \citenamefont {Dussaux}, \citenamefont {Ohashi}, \citenamefont
  {Loretz}, \citenamefont {Schirhagl}, \citenamefont {Watanabe}, \citenamefont
  {Shikata}, \citenamefont {Itoh},\ and\ \citenamefont {Degen}}]{Rosskopf2014}%
  \BibitemOpen
  \bibfield  {author} {\bibinfo {author} {\bibfnamefont {T.}~\bibnamefont
  {Rosskopf}}, \bibinfo {author} {\bibfnamefont {A.}~\bibnamefont {Dussaux}},
  \bibinfo {author} {\bibfnamefont {K.}~\bibnamefont {Ohashi}}, \bibinfo
  {author} {\bibfnamefont {M.}~\bibnamefont {Loretz}}, \bibinfo {author}
  {\bibfnamefont {R.}~\bibnamefont {Schirhagl}}, \bibinfo {author}
  {\bibfnamefont {H.}~\bibnamefont {Watanabe}}, \bibinfo {author}
  {\bibfnamefont {S.}~\bibnamefont {Shikata}}, \bibinfo {author} {\bibfnamefont
  {K.~M.}\ \bibnamefont {Itoh}}, \ and\ \bibinfo {author} {\bibfnamefont
  {C.~L.}\ \bibnamefont {Degen}},\ }\href@noop {} {\bibfield  {journal}
  {\bibinfo  {journal} {Phys. Rev. Lett.}\ }\textbf {\bibinfo {volume} {112}},\
  \bibinfo {pages} {147602} (\bibinfo {year} {2014})}\BibitemShut {NoStop}%
\bibitem [{\citenamefont {Myers}\ \emph {et~al.}(2014)\citenamefont {Myers},
  \citenamefont {Das}, \citenamefont {Dartiailh}, \citenamefont {Ohno},
  \citenamefont {Awschalom},\ and\ \citenamefont
  {Bleszynski~Jayich}}]{Myers2014}%
  \BibitemOpen
  \bibfield  {author} {\bibinfo {author} {\bibfnamefont {B.~A.}\ \bibnamefont
  {Myers}}, \bibinfo {author} {\bibfnamefont {A.}~\bibnamefont {Das}}, \bibinfo
  {author} {\bibfnamefont {M.~C.}\ \bibnamefont {Dartiailh}}, \bibinfo {author}
  {\bibfnamefont {K.}~\bibnamefont {Ohno}}, \bibinfo {author} {\bibfnamefont
  {D.~D.}\ \bibnamefont {Awschalom}}, \ and\ \bibinfo {author} {\bibfnamefont
  {A.~C.}\ \bibnamefont {Bleszynski~Jayich}},\ }\href@noop {} {\bibfield
  {journal} {\bibinfo  {journal} {Phys. Rev. Lett.}\ }\textbf {\bibinfo
  {volume} {113}},\ \bibinfo {pages} {027602} (\bibinfo {year}
  {2014})}\BibitemShut {NoStop}%
\bibitem [{\citenamefont {Romach}\ \emph {et~al.}(2015)\citenamefont {Romach},
  \citenamefont {M\"uller}, \citenamefont {Unden}, \citenamefont {Rogers},
  \citenamefont {Isoda}, \citenamefont {Itoh}, \citenamefont {Markham},
  \citenamefont {Stacey}, \citenamefont {Meijer}, \citenamefont {Pezzagna},
  \citenamefont {Naydenov}, \citenamefont {McGuinness}, \citenamefont
  {Bar-Gill},\ and\ \citenamefont {Jelezko}}]{Romach2015}%
  \BibitemOpen
  \bibfield  {author} {\bibinfo {author} {\bibfnamefont {Y.}~\bibnamefont
  {Romach}}, \bibinfo {author} {\bibfnamefont {C.}~\bibnamefont {M\"uller}},
  \bibinfo {author} {\bibfnamefont {T.}~\bibnamefont {Unden}}, \bibinfo
  {author} {\bibfnamefont {L.~J.}\ \bibnamefont {Rogers}}, \bibinfo {author}
  {\bibfnamefont {T.}~\bibnamefont {Isoda}}, \bibinfo {author} {\bibfnamefont
  {K.~M.}\ \bibnamefont {Itoh}}, \bibinfo {author} {\bibfnamefont
  {M.}~\bibnamefont {Markham}}, \bibinfo {author} {\bibfnamefont
  {A.}~\bibnamefont {Stacey}}, \bibinfo {author} {\bibfnamefont
  {J.}~\bibnamefont {Meijer}}, \bibinfo {author} {\bibfnamefont
  {S.}~\bibnamefont {Pezzagna}}, \bibinfo {author} {\bibfnamefont
  {B.}~\bibnamefont {Naydenov}}, \bibinfo {author} {\bibfnamefont {L.~P.}\
  \bibnamefont {McGuinness}}, \bibinfo {author} {\bibfnamefont
  {N.}~\bibnamefont {Bar-Gill}}, \ and\ \bibinfo {author} {\bibfnamefont
  {F.}~\bibnamefont {Jelezko}},\ }\href@noop {} {\bibfield  {journal} {\bibinfo
   {journal} {Phys. Rev. Lett.}\ }\textbf {\bibinfo {volume} {114}},\ \bibinfo
  {pages} {017601} (\bibinfo {year} {2015})}\BibitemShut {NoStop}%
\bibitem [{\citenamefont {Gordon}\ \emph {et~al.}(2013)\citenamefont {Gordon},
  \citenamefont {Weber}, \citenamefont {Varley}, \citenamefont {Janotti},
  \citenamefont {Awschalom},\ and\ \citenamefont {Van~de Walle}}]{Gordon2013}%
  \BibitemOpen
  \bibfield  {author} {\bibinfo {author} {\bibfnamefont {L.}~\bibnamefont
  {Gordon}}, \bibinfo {author} {\bibfnamefont {J.~R.}\ \bibnamefont {Weber}},
  \bibinfo {author} {\bibfnamefont {J.~B.}\ \bibnamefont {Varley}}, \bibinfo
  {author} {\bibfnamefont {A.}~\bibnamefont {Janotti}}, \bibinfo {author}
  {\bibfnamefont {D.~D.}\ \bibnamefont {Awschalom}}, \ and\ \bibinfo {author}
  {\bibfnamefont {C.~G.}\ \bibnamefont {Van~de Walle}},\ }\href@noop {}
  {\bibfield  {journal} {\bibinfo  {journal} {MRS Bull.}\ }\textbf {\bibinfo
  {volume} {38}},\ \bibinfo {pages} {802} (\bibinfo {year} {2013})}\BibitemShut
  {NoStop}%
\bibitem [{\citenamefont {Ikeda}\ \emph {et~al.}(2010)\citenamefont {Ikeda},
  \citenamefont {Miura}, \citenamefont {Yamamoto}, \citenamefont {Mizunuma},
  \citenamefont {Gan}, \citenamefont {Endo}, \citenamefont {Kanai},
  \citenamefont {Hayakawa}, \citenamefont {Matsukura},\ and\ \citenamefont
  {Ohno}}]{Ikeda2010}%
  \BibitemOpen
  \bibfield  {author} {\bibinfo {author} {\bibfnamefont {S.}~\bibnamefont
  {Ikeda}}, \bibinfo {author} {\bibfnamefont {K.}~\bibnamefont {Miura}},
  \bibinfo {author} {\bibfnamefont {H.}~\bibnamefont {Yamamoto}}, \bibinfo
  {author} {\bibfnamefont {K.}~\bibnamefont {Mizunuma}}, \bibinfo {author}
  {\bibfnamefont {H.~D.}\ \bibnamefont {Gan}}, \bibinfo {author} {\bibfnamefont
  {M.}~\bibnamefont {Endo}}, \bibinfo {author} {\bibfnamefont {S.}~\bibnamefont
  {Kanai}}, \bibinfo {author} {\bibfnamefont {J.}~\bibnamefont {Hayakawa}},
  \bibinfo {author} {\bibfnamefont {F.}~\bibnamefont {Matsukura}}, \ and\
  \bibinfo {author} {\bibfnamefont {H.}~\bibnamefont {Ohno}},\ }\href@noop {}
  {\bibfield  {journal} {\bibinfo  {journal} {Nat. Mater.}\ }\textbf
  {\bibinfo {volume} {9}},\ \bibinfo {pages} {721} (\bibinfo {year}
  {2010})}\BibitemShut {NoStop}%
\bibitem [{\citenamefont {Schwabl}(2004)}]{Schwabl2004}%
  \BibitemOpen
  \bibfield  {author} {\bibinfo {author} {\bibfnamefont {F.}~\bibnamefont
  {Schwabl}},\ }\href@noop {} {\emph {\bibinfo {title} {Advanced Quantum
  Mechanics}}}\ (\bibinfo  {publisher} {Springer}, New York,\ \bibinfo {year}
  {2004})\BibitemShut {NoStop}%
\bibitem [{\citenamefont {Bjorken}\ and\ \citenamefont
  {Drell}(1965)}]{Bjorken1965}%
  \BibitemOpen
  \bibfield  {author} {\bibinfo {author} {\bibfnamefont {J.~D.}\ \bibnamefont
  {Bjorken}}\ and\ \bibinfo {author} {\bibfnamefont {S.~D.}\ \bibnamefont
  {Drell}},\ }\href@noop {} {\emph {\bibinfo {title} {Relativistic Quantum
  Fields}}}\ (\bibinfo  {publisher} {McGraw-Hill}, New York, \ \bibinfo {year}
  {1965})\BibitemShut {NoStop}%
\bibitem [{\citenamefont {Gali}\ \emph {et~al.}(2008)\citenamefont {Gali},
  \citenamefont {Fyta},\ and\ \citenamefont {Kaxiras}}]{Gali2008}%
  \BibitemOpen
  \bibfield  {author} {\bibinfo {author} {\bibfnamefont {A.}~\bibnamefont
  {Gali}}, \bibinfo {author} {\bibfnamefont {M.}~\bibnamefont {Fyta}}, \ and\
  \bibinfo {author} {\bibfnamefont {E.}~\bibnamefont {Kaxiras}},\ }\href@noop
  {} {\bibfield  {journal} {\bibinfo  {journal} {Phys. Rev. B}\ }\textbf
  {\bibinfo {volume} {77}},\ \bibinfo {pages} {155206} (\bibinfo {year}
  {2008})}\BibitemShut {NoStop}%
\end{thebibliography}
\end{document}


\title{Supplemental Material: Atomic-scale magnetometry of dynamic magnetization}

\date{\today}

\author{J. van Bree}
\email{j.v.bree@tue.nl}
\author{M. E. Flatt{\'e}}
\email{michael\_flatte@mailaps.org}
\affiliation{Department of Physics and Astronomy and Optical Science and Technology Center, University of Iowa, Iowa City, Iowa 52242, USA}

\maketitle

\section{Spin density of $N$-particle state}
In the Letter we asserted that $\langle {\bf J}({\bf x}) \rangle = \langle {\bf J} \rangle {\cal P}({\bf x})$. Here we derive this relation for any $N$-particle wave function $\Psi(1,\dots,N)$. In absence of spin-orbit coupling, this wave function is the product of an orbital part $\psi$ and spin part $\chi$,
\begin{align}
\Psi(1,\dots,N) = \psi({\bf x}_1,\dots,{\bf x}_N) \chi(\sigma_1,\dots,\sigma_N),
\end{align}
where ${\bf x}_i$ and $\sigma_i$ are the spatial and spin coordinates of particle $i$. Computing the expectation value of the $N$-particle spin density operator~\cite{Schwabl2004}
\begin{align}
{\bf J}({\bf x}) &= \sum_{i=1}^N \delta({\bf x}-{\bf x}_{i}) {\bf S}_{i},
\end{align}
we obtain the spin density
\begin{align}
\langle {\bf J}({\bf x}) \rangle &= \sum_{i=1}^N \chi(\sigma_1,\dots,\sigma_N)^* {\bf S}_i \chi(\sigma_1,\dots,\sigma_N) \int \psi({\bf x}_1,\dots,{\bf x}_i,\dots,{\bf x}_N)^* \delta({\bf x}-{\bf x}_i) \psi({\bf x}_1,\dots,{\bf x}_i,\dots,{\bf x}_N)~d^3x_1~\dots~d^3x_N \nonumber \\
&= \sum_{i=1}^N \chi(\sigma_1,\dots,\sigma_N)^* {\bf S}_i \chi(\sigma_1,\dots,\sigma_N) \int \psi({\bf x}_i,\dots,{\bf x}_1,\dots,{\bf x}_N)^* \delta({\bf x}-{\bf x}_1) \psi({\bf x}_i,\dots,{\bf x}_1,\dots,{\bf x}_N)~d^3x_1~\dots~d^3x_N \nonumber \\
&= \sum_{i=1}^N \chi(\sigma_1,\dots,\sigma_N)^* {\bf S}_i \chi(\sigma_1,\dots,\sigma_N) \int \psi({\bf x}_1,\dots,{\bf x}_i,\dots,{\bf x}_N)^* \delta({\bf x}-{\bf x}_1) \psi({\bf x}_1,\dots,{\bf x}_i,\dots,{\bf x}_N)~d^3x_1~\dots~d^3x_N \nonumber \\
&= \chi(\sigma_1,\dots,\sigma_N)^* {\bf J} \chi(\sigma_1,\dots,\sigma_N) \int |\psi({\bf x}_1,\dots,{\bf x}_N)|^2 \delta({\bf x}-{\bf x}_1)~d^3x_1~\dots~d^3x_N \nonumber \\
&\equiv \langle {\bf J} \rangle {\cal P}({\bf x}).
\end{align}
In the first step we swapped the integration variables $i$ and $1$. In the second step we used the coordinate permutation properties of $\psi$: since $\psi$ appears twice in the integral, any interchanging of coordinates $1$ and $i$ does not affect the integral. We then recognize that the integral depends no longer on $i$, so the integral can be taken outside of the summation. The spin density depends therefore solely on the total spin operator ${\bf J} = \sum_{i}^N {\bf S}_i$, and is indeed proportional to the spatial probability density ${\cal P}({\bf x})$. We do point out that ${\cal P}({\bf x})$ is generally not a simple summation of the single particle state probabilities: the anti-symmetrisation of $\Psi$ implies that $\psi$ is a linear combination of {\it products} of single particle wave functions, which means ${\cal P}({\bf x})$ depends on the orthogonality of those single particle wave functions.

\section{Magnetic energy Hamiltonian - spherical symmetry}
In the Letter we suggested that the $\eta$-dependence of $E_{\text{mag}}$ is effectively caused by a $J_z^2$ term in spin's ground state Hamiltonian. Here we derive this Hamiltonian term, taking the quantum-mechanical nature of the spin operator into account. As stated in the Letter, a spin's magnetic moment density ${\boldsymbol \mu}({\bf x})$ is generated by a current density ${\bf j}({\bf x}) = \nabla \times {\boldsymbol \mu}({\bf x})$. Whereas in the Letter we used the expectation value of the spin-operator, i.e. assumed the spin is classical, here we recognize that ${\bf j}({\bf x})$ is an operator in the spin-subspace,
\begin{align}
{{\bf j}}({\bf x}) = \frac{2\mu_B}{\hbar} \nabla \times {\cal P}({\bf x}) {{\bf J}} = \frac{2\mu_B}{\hbar} \nabla {\cal P}({\bf x}) \times {{\bf J}},
\end{align}
in absence of spin-orbit interaction. Consider now a slightly more general setup than in the Letter: the spin is localized in region I, adjacent to region II with a different magnetic permeability filling half-space ${\bf x}\cdot{\bf n}\geq 1$. The vector potential in region I is then the sum of ${\bf A}({\bf x})$, due to ${\bf j}({\bf x})$, and $\tilde{{\bf A}}({\bf x})$, due to the image current distribution $\tilde{{\bf j}}({\bf x})$
\begin{align}
\tilde{{\bf j}}({\bf x}) &= \Delta \mu_r \left[{\bf j}(\tilde{\bf x}) - 2 \frac{{\bf j}(\tilde{\bf x}) \cdot {\bf n}}{|{\bf n}|} \frac{\bf n}{|{\bf n}|} \right],
\end{align}
where $\Delta \mu_r = \frac{\mu_r^{\text{II}}-\mu_r^\text{I}}{\mu_r^{\text{II}}+\mu_r^\text{I}}$ is the contrast in magnetic permeability between regions I and II, and $\tilde{\bf x}$ the current image's coordinates
\begin{align}
\tilde{\bf x} = {\bf x} + 2 {\bf n} - 2\frac{{\bf x}\cdot{\bf n}}{|{\bf n}|} \frac{\bf n}{|{\bf n}|}.
\end{align}
Using the relation
\begin{align}
\nabla_{\tilde{\bf x}} f(\tilde{\bf x}) = \nabla_{\bf x} f(\tilde{\bf x}) - 2 \frac{\nabla_{\bf x} f(\tilde{\bf x}) \cdot {\bf n}}{|{\bf n}|} \frac{\bf n}{|\bf n|},
\end{align}
the image current can be expressed as
\begin{align}
\tilde{\bf j}({\bf x}) &= \frac{2\mu_B}{\hbar} \Delta \mu_r \nabla {\cal P}(\tilde{\bf x}) \times \left(\left[2\frac{\bf n}{|{\bf n}|} \cdot {\bf J}\right] \frac{\bf n}{|{\bf n}|} - {\bf J} \right)
\end{align}
using vector calculus; we take $\nabla$ to act on ${\bf x}$ unless indicated differently. Expressing the vector potential in terms of the current density
\begin{align}
{\bf A}({\bf x}) = \frac{\mu_0}{4 \pi} \int_V \frac{{\bf j}({\bf x'})}{|{\bf x}-{\bf x}'|}~d^3x',
\end{align}
we can write, in the Coulomb gauge, the EM-field Hamiltonian ${\cal H}_{\text{mag}}$ in the spin subspace as~\cite{Bjorken1965,Jackson1998}
\begin{align}
{\cal H}_{\text{mag}} &= \frac{1}{2} \int_V \left[{\bf A}({\bf x})+\tilde{{\bf A}}({\bf x})\right] \cdot {\bf j}({\bf x})~d^3x \\
&= \frac{\mu_0 \mu_r^{\text{I}} \mu_B^2}{2\pi \hbar^2} \int_{V} \left[\nabla \zeta({\bf x},{\bf 0}) \times {\bf J}\right] \cdot \left[\nabla {\cal P}({\bf x}) \times {\bf J}\right]~d^3x \nonumber \\
&\hspace{0.5cm} + \frac{\mu_0 \mu_r^{\text{I}} \mu_B^2}{2\pi \hbar^2} \Delta \mu_r \int_{V} \left[\nabla \zeta({\bf x},{\bf n}) \times \left(\left[2\frac{\bf n}{|{\bf n}|} \cdot {\bf J}\right] \frac{\bf n}{|{\bf n}|} - {\bf J} \right)\right] \cdot \left[\nabla {\cal P}(\bf x) \times {\bf J}\right]~d^3x. \label{eq:H}
\end{align}
Here we have defined $\zeta({\bf x},{\bf n})$ through the relation
\begin{align}
\int_{\tilde V} \frac{\nabla' {\cal P}(\tilde{\bf x}')}{|{\bf x}-{\bf x}'|}~d^3x' &= \left. \frac{{\cal P}(\tilde{\bf x}')}{|{\bf x}-{\bf x}'|} \right|_{{\bf x}' \in \tilde{S}} - \int_{\tilde{V}} {\cal P}(\tilde{\bf x}')\nabla' \left(\frac{1}{|{\bf x}-{\bf x}'|}\right)~d^3x' \\
&= \nabla \left(\int_{\tilde{V}} \frac{{\cal P}(\tilde{\bf x}')}{|{\bf x}-{\bf x}'|}~d^3x'\right) \\
&= \nabla \left(\int_{V} \frac{{\cal P}({\bf x}')}{\left|{\bf x}-\left({\bf x}'+2{\bf n}-2\frac{{\bf x}'\cdot{\bf n}}{|{\bf n}|} \frac{\bf n}{|{\bf n}|}\right)\right|}~d^3x'\right) \\
&\equiv \nabla \zeta({\bf x},{\bf n}),
\end{align}
and assumed the spin to be localized in region I, so that integration volume $V$ ($\tilde{V}$) is limited to region I (II) and ${\cal P}({\bf x})=0$ on their surfaces. By defining ${\bf x}_{\bf n}'={\bf x}'+2{\bf n}-2\frac{{\bf x}'\cdot{\bf n}}{|{\bf n}|} \frac{\bf n}{|{\bf n}|}$, the nominator in $\zeta({\bf x},{\bf n})$ can be expanded in spherical harmonics $Y_l^m(\theta,\phi)$,
\begin{align}
\zeta({\bf x},{\bf n}) &= \sum_{l=0}^{\infty} \sum_{m=-l}^{l} \frac{4\pi}{2l+1} Y_l^m(\theta,\phi) \int_{V} {\cal P}({\bf x}') \frac{r_<^l}{r_>^{l+1}} Y_l^{m}({\theta}_{\bf n}',{\phi}_{\bf n}')^*~d^3x' \equiv \sum_{l=0}^{\infty} \sum_{m=-l}^{l} \frac{4\pi}{2l+1} Y_l^m(\theta,\phi) {\cal R}_l^m(r,{\bf n}),
\end{align}
where $r_<$ ($r_>$) is the smaller (larger) of $r$ and $r_{\bf n}'$. We now revert to the setup of the Letter, i.e. ${\bf n}=d~{\bf e}_z$, so that
\begin{align}
r_{\bf n}' &= \sqrt{4d^2+r'^2-4dr'\cos\theta'} \\
\cos {\theta}_{\bf n}' &= \frac{2d - r' \cos \theta'}{\sqrt{4d^2+r'^2-4dr'\cos\theta'}} \\
{\phi}_{\bf n}' &= \phi',
\end{align}
and assume spherical symmetry, ${\cal P}({\bf x})={\cal P}(r)$. It is straightforward to see that ${\cal R}_l^m(r,{\bf n})$ is only non-zero for $m=0$, and ${\cal R}_l^m(r,{\bf 0})$ is only non-zero for $l=0$. Therefore
\begin{align}
{\cal H}_{\text{mag}}^{\text{sph}} &= \frac{\mu_0 \mu_r^{\text{I}} \mu_B^2}{\sqrt{\pi} \hbar^2} \int_{V} \left[\nabla {\cal R}_0^0(r,{\bf 0}) \times {\bf J}\right] \cdot \left[\nabla {\cal P}(r) \times {\bf J}\right]~d^3x \nonumber \\
&\hspace{0.5cm} + \sum_{l=0}^{\infty} \frac{2 \mu_0 \mu_r^{\text{I}} \mu_B^2}{(2l+1)\hbar^2} \Delta \mu_r \int_{V} \left[\nabla \left\{Y_l^0(\theta,\phi) {\cal R}_l^0(r,{\bf n})\right\} \times \left(2 J_z {\bf e}_z - {\bf J} \right)\right] \cdot \left[\nabla {\cal P}(r) \times {\bf J}\right]~d^3x.
\end{align}
Integrating over $\phi$ and $\theta$ results in
\begin{align}
{\cal H}_{\text{mag}}^{\text{sph}} &= \frac{8 \sqrt{\pi}\mu_0 \mu_r^{\text{I}} \mu_B^2}{3 \hbar^2} \int \frac{d{\cal R}_0^0(r,{\bf 0})}{dr} \frac{d{\cal P}(r)}{dr} r^2~dr \left(J_x^2+J_y^2+J_z^2\right) \nonumber \\
&\hspace{0.5cm} - \frac{4 \sqrt{\pi}\mu_0 \mu_r^{\text{I}} \mu_B^2}{15 \sqrt{5} \hbar^2} \Delta \mu_r \int \left(\frac{3}{r} {\cal R}_2^0(r,{\bf n}) + \frac{d{\cal R}_2^0(r,{\bf n})}{dr} + 10 \sqrt{5} \frac{d{\cal R}_0^0(r,{\bf n})}{dr}\right) \frac{d{\cal P}(r)}{dr} r^2~dr \left(J_x^2+J_y^2+J_z^2\right) \nonumber \\
&\hspace{0.5cm} - \frac{4 \sqrt{\pi}\mu_0 \mu_r^{\text{I}} \mu_B^2}{15 \sqrt{5} \hbar^2} \Delta \mu_r \int \left(\frac{3}{r} {\cal R}_2^0(r,{\bf n}) + \frac{d{\cal R}_2^0(r,{\bf n})}{dr} - 20 \sqrt{5} \frac{d{\cal R}_0^0(r,{\bf n})}{dr}\right) \frac{d{\cal P}(r)}{dr} r^2~dr~J_z^2. \label{eq:sph}
\end{align}
Only spherical harmonics with $l=0$ and $l=2$ appear in the part of the Hamiltonian related to region II. As we will show explicitly later on, the $l=0$ term vanishes after radial integration. The $l=2$ contribution does not vanish and its origin can be explained as follows. The Hamiltonian describes the interaction between the spin's magnetic moment density and the magnetic field it induces in region II. Expanding these in terms of spherical harmonics, they have respectively $l=0$ and $l=2$ in case of spherical symmetry. Their product enters the Hamiltonian, which is proportional to a single spherical harmonic of order $l=0+2$. Therefore only of the $l=2$ term in Eq.~\ref{eq:sph} will lead to a contribution to the Hamiltonian. Indeed, if we work out the integrals over $r$, we observe that
\begin{align}
{\cal R}_0^0(r,{\bf 0}) &= 2\sqrt{\pi} \left[\frac{1}{r} \int_0^r {\cal P}(r') r'^2~dr' + \int_r^R {\cal P}(r') r'~dr'\right],
\end{align}
so that 
\begin{align}
\int \frac{d{\cal R}^0_0(r,{\bf 0})}{dr} \frac{d {\cal P}(r)}{dr} r^2~dr &= 2\sqrt{\pi} \int_0^R {\cal P}(r)^2 r^2~dr,
\end{align}
where, in line with the spin being localized in region I, ${\cal P}(R)=0$ has been assumed. We also find
\begin{align}
{\cal R}_l^0(r,{\bf n}) &= \sqrt{(2l+1)\pi} r^l \int {\cal P}(r') r'^2 \left(\int \frac{P_l^0(\cos \theta_{\bf n}')}{r_{\bf n}'^{l+1}}\sin\theta'~d\theta'\right)dr' = \sqrt{\frac{(2l+1)}{16\pi}} \frac{r^l}{2^l d^{l+1}},
\end{align}
where we used $\int_0^R {\cal P}(r) r^2~dr = \tfrac{1}{4\pi}$. Therefore
\begin{align}
\int \frac{d{\cal R}_l^0(r,{\bf n})}{dr} \frac{d {\cal P}(r)}{dr} r^2~dr &= -\sqrt{\frac{(2l+1)}{16\pi}} \frac{l(l+1)}{2^l d^{l+1}} \int_0^R {\cal P}(r) r^l~dr \\
\int {\cal R}_l^0(r,{\bf n}) \frac{d {\cal P}(r)}{dr} r~dr &= -\sqrt{\frac{(2l+1)}{16\pi}} \frac{l+1}{2^l d^{l+1}} \int_0^R {\cal P}(r) r^l~dr,
\end{align}
where we explicitly see that the $l=0$ contribution vanishes after radial integration. Using these relations, the Hamiltonian becomes
\begin{align}
{\cal H}_{\text{mag}}^{\text{sph}} &= \frac{\mu_0 \mu_r^{\text{I}} \mu_B^2}{\hbar^2} \left[\frac{16 \pi}{3} \int_0^R {\cal P}(r)^2 r^2~dr + \frac{\Delta \mu_r}{16 \pi d^3} \right] \left(J_x^2+J_y^2+J_z^2\right) + \frac{\mu_0 \mu_r^{\text{I}} \mu_B^2}{16 \pi \hbar^2 d^3} \Delta \mu_r J_z^2.
\end{align}
Indeed we find that the spin's magnetic energy leads to additional terms in the Hamiltonian. In particular, the introduction of region II leads to an effective Hamiltonian ${\cal H}_{\text{eff}} = D_{\text{mag}}^{\text{sph}} J_z^2$, where
\begin{align}
D_{\text{mag}}^{\text{sph}} = \left(\frac{\mu_r^{\text{II}}-\mu_r^\text{I}}{\mu_r^{\text{II}}+\mu_r^\text{I}}\right) \frac{\mu_0 \mu_r^{\text{I}} \mu_B^2}{16 \pi \hbar^2 d^3}.
\end{align}
As their pre-factors differ, it might seem ${\cal H}_{\text{mag}}^{\text{sph}}$ and $E_{\text{mag}}$ are incompatible. However, this difference arises from the classical treatment of the spin for $E_{\text{mag}}$ versus the quantum-mechanical treatment for ${\cal H}_{\text{mag}}$. To make a direct comparison, we need to calculate the magnetic energy resulting from ${\cal H}_{\text{mag}}^{\text{sph}}$ for a spin making an angle $\eta$ with respect to the $z$-axis. It is more convenient to use a basis rotated by the angle $\eta$, so that the spin is in the state $|J,J\rangle$ and ${\cal H}_{\text{mag}}^{\text{sph}}$ becomes
 \begin{align}
{\cal H}_{\text{mag}}^{\text{sph}} &= \frac{\mu_0 \mu_r^{\text{I}} \mu_B^2}{\hbar^2} \left[\frac{16 \pi}{3} \int_0^R {\cal P}(r)^2 r^2~dr + \frac{\Delta \mu_r}{16 \pi d^3} \right] \left(J_x^2+J_y^2+J_z^2\right) + \frac{\mu_0 \mu_r^{\text{I}} \mu_B^2}{16 \pi \hbar^2 d^3} \Delta \mu_r \left(J_x \sin \eta + J_z \cos \eta\right)^2.
\end{align}
The magnetic energy of this state is then
\begin{align}
\langle J,J | {\cal H}_{\text{mag}}^{\text{sph}} | J,J \rangle = \frac{16}{3} \mu_0 \mu_r^{\text{I}} \mu_B^2 J^2 \pi \int_0^R {\cal P}(r)^2 r^2~dr + \frac{\mu_0 \mu_r^{\text{I}} \mu_B^2 J^2}{32 \pi d^3} \Delta \mu_r \left(3 + \cos 2\eta + \frac{1}{2J} \left[1 - \cos 2\eta\right]\right).
\end{align}
Hence we find $\lim_{J \rightarrow \infty} \langle J,J | {\cal H}_{\text{mag}}^{\text{sph}} | J,J \rangle = E_{\text{mag}}$, as one would expect in the semiclassical (large-J) limit of a quantum system.

\section{Magnetic energy Hamiltonian - cylindrical symmetry}
\begin{figure*}
\includegraphics[width=\textwidth]{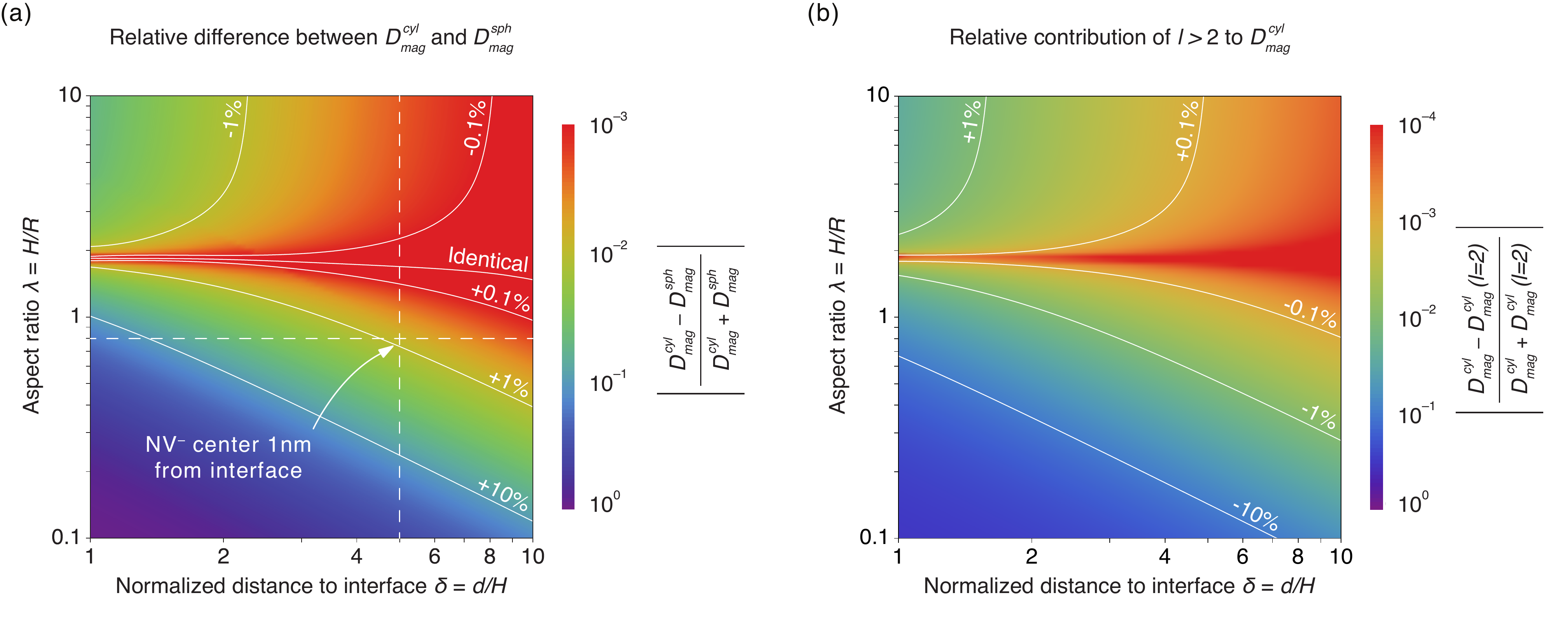}
\caption{(a) The relative difference between $D_{\text{mag}}$ with cylindrical and spherical symmetry, as a function of the spin's normalized distance to the interface $\delta = d/H$ and its aspect ratio $\lambda = H/R$. The continuous white lines are contours of constant relative difference; $D_{\text{mag}}^{\text{cyl}} = D_{\text{mag}}^{\text{sph}}$ along the line with label 'identical'. For $H > R$ or large enough $\delta$ the difference is less than $5\%$. The dashed lines indicate the estimated NV$^-$ center's aspect ratio (horizontal) and when it is 1~nm away from the interface to region II (vertical). In that case $D_{\text{mag}}^{\text{cyl}}$ deviates less than $1\%$ from $D_{\text{mag}}^{\text{sph}}$. (b) The relative contribution of the $l>2$ terms to $D_{\text{mag}}^{\text{cyl}}$ as function of $\delta$ and $\lambda$. For $H>R$ or large enough $\delta$, the $l>2$ terms are contributing less than $1\%$ to $D_{\text{mag}}^{\text{cyl}}$.}
\label{fig:DCylvsSph}
\end{figure*}

In the Letter we calculated the NV$^-$ center's magnetic energy assuming its probability density ${\cal P}({\bf x})$ has spherical symmetry, instead of its actual $C_{3v}$-symmetry~\cite{Doherty2013}. Here we will validate this approximation. We start by calculating ${\cal H}_{\text{mag}}$ assuming cylindrical $D_{\infty h}$-symmetry, ${\cal P}({\bf x})={\cal P}(\rho,z)$. We take the axial symmetry axis of ${\cal P}(\rho,z)$ perpendicular to the interface between the two regions. It has been demonstrated that such orientation can be realized deterministically in practice~\cite{Michl2014}. Following the same setup as described in the Letter, i.e. ${\bf n}=d~{\bf e}_z$, ${\cal H}_{\text{mag}}$ (Eq.~\ref{eq:H}) is now
\begin{align}
{\cal H}_{\text{mag}}^{\text{cyl}} &= \sum_{l=0}^{\infty} \frac{2 \mu_0 \mu_r^{\text{I}} \mu_B^2}{(2l+1)\hbar^2} \int_{V} \left[\nabla \left\{Y_l^0(\theta,\phi) {\cal R}_l^0(r,{\bf 0})\right\} \times {\bf J}\right] \cdot \left[\nabla {\cal P}(\rho,z) \times {\bf J}\right]~d^3x \nonumber \\
&\hspace{0.5cm} + \sum_{l=0}^{\infty} \frac{2 \mu_0 \mu_r^{\text{I}} \mu_B^2}{(2l+1)\hbar^2} \Delta \mu_r \int_{V} \left[\nabla \left\{Y_l^0(\theta,\phi) {\cal R}_l^0(r,{\bf n})\right\} \times \left(2 J_z {\bf e}_z - {\bf J} \right)\right] \cdot \left[\nabla {\cal P}(\rho,z) \times {\bf J}\right]~d^3x,
\end{align}
since ${\cal R}_l^m(r,{\bf n})$ is only non-zero for $m=0$. The relation
\begin{align}
\iint \frac{\partial \left\{Y_l^0(\theta,\phi) {\cal R}_l^0(r,{\bf n}) \right\}}{\partial \rho} \frac{\partial {\cal P}(\rho,z)}{\partial \rho} \rho~d\rho dz = -\iint \frac{\partial \left\{Y_l^0(\theta,\phi) {\cal R}_l^0(r,{\bf n}) \right\}}{\partial z} \frac{\partial {\cal P}(\rho,z)}{\partial z} \rho~d\rho dz \equiv M_l({\bf n}),
\end{align}
can be derived using partial integration and the property $\nabla^2 \left\{Y_l^0(\theta,\phi) {\cal R}_l^0(r,{\bf n}) \right\} = 0$ (valid by construction). Using this relation and integrating over $\phi$, we obtain
\begin{align}
{\cal H}_{\text{mag}}^{\text{cyl}} &= \frac{2 \pi \mu_0 \mu_r^{\text{I}} \mu_B^2}{\hbar^2} \sum_{l=0}^{\infty} \frac{\Delta \mu_r M_l({\bf n}) - M_l({\bf 0})}{(2l+1)} \left(J_x^2+J_y^2+J_z^2\right) + \frac{2 \pi \mu_0 \mu_r^{\text{I}} \mu_B^2}{\hbar^2} \sum_{l=0}^{\infty} \frac{\Delta \mu_r M_l({\bf n}) + 3 M_l({\bf 0})}{(2l+1)} J_z^2.
\end{align}
Region II leads again to a $J_z^2$-term in the Hamiltonian: lowering the symmetry of ${\cal P}({\bf x})$ to $D_{\infty h}$ does not alter the structure of the Hamiltonian. The $J_z^2$-term is also present when region II is absent, which is caused by the $D_{\infty h}$-symmetry ${\cal P}({\bf x})$. This term is irrelevant, however, as it cannot be distinguished from other mechanisms contributing to the fine structure constant $D_{\text{GS}}$. (If this were possible, it would provide a measure of the aspect ratio of ${\cal P}({\bf x})$.) To quantitatively compare ${\cal H}_{\text{mag}}^{\text{cyl}}$ with ${\cal H}_{\text{mag}}^{\text{sph}}$, we take the spin confined to a cylinder with radius $R$ and height $H$,
\begin{align}
{\cal P}(\rho,z) = N \left[J_0\left(\frac{\rho_{0,1}}{R} \rho\right) \cos \left(\frac{\pi}{H}z\right)\right]^2,
\end{align}
where $N$ is a normalisation constant, and $\rho_{0,1}$ the first zero of the $0^{\text{th}}$-order Bessel function $J_0(x)$. Concentrating on the $J_z^2$ term in the Hamiltonian caused by the presence of region II, we find it can be expressed as
\begin{align}
\frac{\mu_0 \mu_r^{\text{I}}\mu_B^2}{16 \pi \hbar^2 H R^2} \Delta \mu_r \sum_{l=2}^{\infty} f_{2l-2}(\delta,\lambda) J_z^2 = D_{\text{mag}}^{\text{cyl}} J_z^2, \label{eq:DCyl}
\end{align}
where $\delta = d/H$ is the normalized distance of the spin to the interface between regions I and II, $\lambda = H/R$ is the aspect ratio. The numbers $f$ depend only on $\delta$ and $\lambda$, are only non-zero for even $l$ due to the even symmetry of ${\cal P}({\bf x})$ in $z$, and start from $l=2$ for the same reasons as for spherical symmetry (see discussion below Eq.~\ref{eq:sph}). Interestingly, the relative difference
\begin{align}
\frac{D_{\text{mag}}^{\text{cyl}} - D_{\text{mag}}^{\text{sph}}}{D_{\text{mag}}^{\text{cyl}} + D_{\text{mag}}^{\text{sph}}} = \frac{\delta^3 \lambda^2 \sum_{l=2}^{\infty} f_l(\delta,\lambda) - 1}{\delta^3 \lambda^2 \sum_{l=2}^{\infty} f_l(\delta,\lambda) +1},
\end{align}
depends only on $\delta$ and $\lambda$. From Fig.~\ref{fig:DCylvsSph}(a) it is clear that for either $H > R$ or large enough $\delta$, the difference between spherical and cylindrical symmetry is $\leq 5\%$: in both cases the cylindrical ${\cal P}({\bf x})$ appears as approximately spherical from region II's perspective. The $l=2$ term is almost entirely responsible for this difference, as can be seen from Fig.~\ref{fig:DCylvsSph}(b). Indeed, the numbers $f$ in Eq.~\ref{eq:DCyl} depend on the distance as $1/\delta^{l+1}$, so that at fixed distance terms with increasing $l$ become less important.

Comparing the NV$^-$ center's $C_{3v}$-symmetry with cylindrical $D_{\infty h}$-symmetry, the probability density is no longer symmetric in $z$ and has three-fold rotation symmetry around the $z$-axis. This reduction in symmetry allows terms proportional to spherical harmonics with $l=2,3,4,...$ and $m=0,\pm3$ to contribute to $D_{\text{mag}}$. We see that the first correction with respect to $D_{\infty h}$-symmetry has to come from the $l=3$ term. It is $1/\delta$ smaller than the $l=2$ term. Based on DFT calculations of the spin density~\cite{Gali2008}, we estimate the NV$^-$-center ground state has $H\approx 2.0$~\AA~and $R\approx 2.5$~\AA, hence $\lambda \approx 0.8$. For $d>1$~nm ($\delta>5$), the corrections of the $l=3$ term are $<20\%$ compared to the $l=2$ term. It is therefore reasonable to approximate the NV$^-$ center's $C_{3v}$-symmetry with cylindrical symmetry. From Fig.~\ref{fig:DCylvsSph}(a) we see that the deviations of $D_{\text{mag}}^{\text{cyl}}$ from $D_{\text{mag}}^{\text{sph}}$ are less than $1\%$. We therefore conclude that taking the NV$^-$ center's wave function as spherically symmetric is a good approximation.